\documentclass{imsart}

\RequirePackage[OT1]{fontenc}
\RequirePackage{amsthm,amsmath}
\RequirePackage[numbers]{natbib}
\usepackage{hyperref}
\hypersetup{colorlinks,citecolor=blue,urlcolor=blue}

\usepackage{graphicx,psfrag,epsf}
\usepackage{enumerate}
\usepackage{url} 

\usepackage{rotating,pdflscape,longtable}

\startlocaldefs
\numberwithin{equation}{section}
\theoremstyle{plain}
\newtheorem{thm}{Theorem}[section]
\endlocaldefs

\begin{document}

\begin{frontmatter}
\title{Practical Valid Inferences for the Two-Sample Binomial Problem}
\runtitle{Two-Sample Binomial Problem}

\begin{aug}
\author{\fnms{Michael P.} \snm{Fay}\ead[label=e1]{mfay@niaid.nih.gov}}
\and
\author{\fnms{Sally A.} \snm{Hunsberger}\ead[label=e2]{Sally.Hunsberger@nih.gov}}

\address{Biostatistics Research Branch \\
National Institute of Allergy and Infectious Diseases \\
Bethesda, MD, USA \\
\printead{e1,e2}}

\runauthor{Fay and Hunsberger}

\affiliation{National Institute of Allergy and Infectious Diseases}

\end{aug}

\begin{abstract}
 Our interest is whether two binomial parameters  differ, which parameter is larger, and by how much. This apparently simple problem was addressed by Fisher in the 1930's, and has been the subject of many review papers since then. Yet there continues to be new work on this issue and no consensus solution. Previous reviews have focused primarily on testing and the properties of validity and power, or primarily on confidence intervals, their coverage, and expected length. Here we evaluate both. For example, we consider whether a p-value and its matching confidence interval are compatible, meaning that the p-value rejects at level $\alpha$ if and only if the $1-\alpha$ confidence interval excludes all null parameter values. For focus, we only examine  non-asymptotic inferences, so that most of the p-values and confidence intervals are valid (i.e., exact) by construction. Within this focus, we review different methods emphasizing many of the properties and interpretational aspects we desire from applied frequentist inference: validity, accuracy, good power, equivariance, compatibility, coherence, and parameterization and direction of effect. We show that no one method can meet all the desirable properties and give recommendations based on which properties are given more importance.
\end{abstract}


\tableofcontents
\end{frontmatter}

\section{Introduction}


Suppose we observe two independent binomial variates
with parameters $(n_1,\theta_1)$ and $(n_2,\theta_2)$.
Two questions are: are $\theta_1$ and $\theta_2$ equal? and
how much larger is one $\theta$ parameter than the other?
To answer these two questions, the frequentist typically presents an estimate of an effect,
a confidence interval (CI) on that effect, and a p-value to test that there is no effect.
Surprisingly, there is no consensus method for testing and creating confidence intervals for this problem.
New methods continue to be developed for this problem \citep[see e.g.,][]{Lloy:2008,Wang:2010,Wang:2015,FayP:2015,Gabr:2018}.
Many review papers focus on testing alone \citep[see][]{Lyde:2009,Ripa:2017}, or confidence intervals alone
\citep[see][]{Newc:1998,Sant:2007,Fage:2015}.
Here we focus on both.

We limit the scope of this paper by considering only frequentist approaches (so Bayesian methods are not covered), and by  not considering asymptotic methods or other approximations.
Many review papers or books
\citep[see e.g.,][]{Newc:1998, Lyde:2009, Fage:2015,Newc:2013}
cover and compare many of those approximations.
Sometimes those  approximations are closed-form expressions and can be useful for deriving simple sample size formulas or when the
test is applied many times such as in genomics. But often the approximations are unnecessary with modern computers.
Non-asymptotic methods are often called exact,
but in this paper we  reserve the term exact for non-asymptotic methods that are valid,
meaning tests that control the type I error rate, and confidence intervals
that cover the parameter with at least the nominal value. See Section~\ref{sec-background} for further discussion of the term exact.
A class of important non-asymptotic tests that are not valid are  mid-p methods (Section~\ref{sec-midp}), which are sometimes called quasi-exact \citep{Hirj:1991} and are included in our review because, for confidence intervals, sometimes we want average coverage close to the nominal value instead of guaranteed coverage that on average is conservative.

Here is an outline of our paper.
Section~\ref{sec-NormalIntuition} begins  by contrasting the two-sample binomial problem with the
two-sample difference in normal distributions with the same variance, in which there is an accepted solution: the two-sample
t-test. This allows us to define inferential properties of interest as well as highlight why there is no single accepted solution to the two-sample binomial problem.
Newcombe \citep{Newc:2013} takes a similar approach.
Section~\ref{sec-EffectChoice}  discusses the choice of effect measure
(e.g., difference in binomial parameters, ratio of parameters,
or odds ratio of parameters).
Section~\ref{sec-methods}  defines a frequentist triple as a parameter estimator,
an associated confidence interval procedure, and a p-value function.
We then formally discuss some properties of triples, such as whether the confidence interval
and p-value match and are compatible, and whether directional inferences may be made from the triple.
The idea of matched triples is discussed in Hirji \citep[][p. 77]{Hirj:2006} in a less formal way as a ``unified report''.
Our review says very little about parameter estimators, and  mostly focuses on properties of p-values and confidence intervals
and the compatibility of p-values with confidence intervals.
Our discussion of directional inferences is motivated from the three decision rule of Neyman \citep[see e.g.,][]{Free:2008}.
We describe methods for defining valid one-sided decision rules in Sections~\ref{sec-OnesidedUncond} (unconditional methods)
and \ref{sec-oneSidedCond} (conditional methods), including the associated p-values and confidence intervals.
Much of Sections~\ref{sec-OnesidedUncond} and \ref{sec-oneSidedCond} was thoroughly reviewed in \citep{Ripa:2017}
but is included in this paper for completeness; however, Section~\ref{sec-orderingRatio} presents some new ideas on informativeness of ordering functions.
Section~\ref{sec-Melded} reviews the melded confidence intervals of \citet{FayP:2015} which
are compatible with the one-sided conditional method (i.e., Fisher's exact test)  p-values.
Section~\ref{sec-twoSidedTests} discusses non-central confidence intervals and associated tests, with a new focus on the relationship of these intervals to directional inferences.
Section~\ref{sec-midp} discusses mid-p methods, which are non-asymptotic methods that relax the validity assumption in order to achieve better accuracy.
Section~\ref{sec-Computations} discusses the computational aspects of  various methods.
Section~\ref{sec-power} discusses power and efficiency of methods,
including some new calculations.
Section~\ref{sec-Recommendations} presents recommendations.
Briefly, we recommend using exact central confidence intervals (those with equal error bounds on both sides) because it is better for directional inferences.
For fast calculations, use exact conditional tests with compatible confidence intervals, but for more power consider exact unconditional test using the version that orders by the
one-sided mid-p Fisher's exact p-values.  If validity is not vital, use mid-p values on the exact conditional test, which are often a good approximation to the exact unconditional tests.

\section{Overview: Failure of Normal Intuition  }
\label{sec-NormalIntuition}

\subsection{Frequentist Inferences}

We define a frequentist triple (or just a triple) as an estimator of a parameter of interest, a confidence interval, and a p-value function.
This approach allows us to compare different triples by examining
 not just properties of each component (i.e., comparing powers of different p-value functions or expected lengths of different confidence intervals), but also to examine properties of the triples as a whole.
For example, within a triple, we examine inferential agreement between the p-value function and confidence interval procedure.
Additionally, we examine what directional inferential statements we can make from the triple,
such as whether $\theta_2$ is significantly larger than $\theta_1$, and at what significance level.

Although in some different statistical settings (e.g., two-sample normal problem) the standard triple will automatically give
inferential agreement between p-values and confidence intervals as well as automatically give directional inferential statements,
in the two-sample binomial problem those inferential properties are not automatic.
Thus, before discussing the binomial problem,
 we  review the two-sample problem with normally distributed responses
with the same variance.
We consider the latter problem first, because there is some consensus that one triple (the difference in means, and the confidence interval and p-value associated with the t-test)
 is appropriate for this problem. In the normal case, this t-test triple
 meets some regularity properties that lead to inferences that are intuitive and easy to understand.
Because these properties form the basis for a certain statistical intuition about how frequentist inferences ought to be,
and because the example uses normal distributional assumptions, we call these properties the ``normal intuition''.
We show later how the normal intuition breaks down for the two-sample binomial problem,
although many of the properties may approximately hold for large samples.

\subsection{Background and Notation}
\label{sec-background}

Consider a general frequentist problem, where we observe data, ${\bf x}$, and denote its random variable as ${\bf X}$.
Assume some probability model for ${\bf X}$ that depends on a parameter vector $\theta$, but we are interested in a function
of $\theta$ that returns a scalar, $b(\theta) = \beta$.
We partition the possible values of $\theta$ into two sets, the null hypothesis space, $\Theta_0$, and the alternative hypothesis space, $\Theta_1$.

In this paper,
 we consider only three classes of partitions, where the null and alternative space is defined by $\beta$, and separated by a
value $\beta_0$ on the boundary between the null and alternative hypothesis spaces.
The first of these three classes are two-sided hypotheses,
\begin{eqnarray*}
H_{0}: & & \beta=\beta_0 \\
H_{1}: & & \beta \neq \beta_0
\end{eqnarray*}
which can be equivalently written
as
\begin{eqnarray*}
H_{0}: & & \theta \in \Theta_0    \mbox{ where }  \Theta_0 = \left\{ \theta : b(\theta) = \beta_0 \right\}  \\
H_{1}: & & \theta \in \Theta_1  \mbox{ where }  \Theta_1 = \left\{ \theta : b(\theta) \neq \beta_0 \right\}.
\end{eqnarray*}
The other two classes are the one-sided hypotheses,
\begin{eqnarray*}
& \mbox{\underline{Alternative is Less}} & \mbox{\underline{Alternative is Greater}} \\
& H_{0}: \beta \geq \beta_0 & H_{0}: \beta \leq \beta_0 \\
& H_{1}: \beta < \beta_0 & H_{1}: \beta > \beta_0.
\end{eqnarray*}

Let  $p({\bf x}, \Theta_0)$ be a p-value associated with the null hypothesis space, $\Theta_0$.
Typically, we assume a class of hypotheses and write (with a slight abuse of notation)
 $p({\bf x}, \beta_0)$ as a p-value associated with the null hypothesis indexed by $\beta_0$.
We reject the null hypothesis at level $\alpha$ if $p({\bf x}, \beta_0) \leq \alpha$.
Following \citet{Berg:1994}, we define a p-value procedure as {\it valid}  if
\begin{eqnarray*}
P_{\theta} \left[ p({\bf X}, \beta_0) \right.  \leq  \left. \alpha  \right] & \leq & \alpha,
\end{eqnarray*}
for all $\alpha \in (0,1)$ and all $\theta \in \Theta_0$.
(Ripamonti, {\it et al} \cite{Ripa:2017} call a valid p-value procedure a {\it guaranteed} p-value.)
The term {\it exact} is often used to describe tests that give valid p-values,
but be aware that the term `exact' is used in at least 4 different ways in the literature: (i)   methods not based on
asymptotic or other approximations  \citep[see][p.450]{Hirj:2006}, (ii)
 valid methods
\citep[see][]{Hirj:1991, Lyde:2009, FayP:2015}, (iii) methods where  the size is equal to the significance level (only possible with randomized tests for discrete data)
\citep{Dude:2007}, or (iv) methods where the p-values are the smallest p-values among a class of valid p-values \citep[][equation 2.5]{Ripa:2017},
specifically, p-value procedures such that
\begin{eqnarray}
\sup_{\theta \in \Theta_0} P_{\theta} \left[ p({\bf X}, \beta_0) \right.  \leq  \left. p({\bf x}, \beta_0) \right] = p({\bf x}, \beta_0) \mbox{ for all {\bf x}},  \label{eq:RipaExact}
\end{eqnarray}
In this review, we use the term exact only in the sense of (ii).

Following \citet{Rohm:2005}, we
define a p-value procedure as {\it coherent}
if  for every ${\bf x}$, $p({\bf x}, \Theta_0^*) \leq p({\bf x}, \Theta_0)$
if $\Theta_0^* \subseteq \Theta_0$.

For the classes of hypotheses above, we can invert the p-value function to get its associated $100(1-\alpha)\%$  confidence region,
 \begin{eqnarray}
C({\bf x}, 1-\alpha) & = & \left\{ \beta: p({\bf x}, \beta) > \alpha \right\}. \label{eq:CIbyp}
\end{eqnarray}
We define a confidence region as {\it valid} if it is guaranteed to have at least nominal coverage  for every $\theta$ (and hence every  $b(\theta)=\beta$);
in other words,
\begin{eqnarray*}
P_{\theta} \left[\beta  \in C({\bf X}, 1-\alpha)  \right] & \geq & 1- \alpha.
\end{eqnarray*}

This paper considers non-asymptotic methods, and all are valid except the mid-p methods described in Section~\ref{sec-midp}.

\subsection{Standard Frequentist Inference: Normal Intuition}


Consider the two-sample problem, where the $a$th group has $n_a$  independent and  normally distributed responses, with mean $\mu_a$ and variance $\sigma^2$, for $a=1,2$.
Let $\theta = [\mu_1, \mu_2, \sigma]$, and suppose we are interested in $\beta = b(\theta) = \mu_2-\mu_1$.
The t-test is valid for testing the null that $\beta=\beta_0$ and it is the uniformly most powerful (UMP) unbiased  test \citep[][p. 160]{Lehm:2005}.
UMP unbiasedness means that among the class of unbiased tests  (i.e., tests for which power for each  parameter value in the alternative space is at least as large as
the  power for every parameter value in the null space),  the t-test is the most powerful test for each $\theta \in \Theta_1$.

We study this case first to define ``normal intuition'' about frequentist inferences.
This normal intuition is a series of properties, that if they are not met, conflict with
many statisticians' intuitive feeling of how p-values and confidence regions ought to work.
Here are those properties met by the triple: difference in sample means, $\hat{\beta}$; the two-sided p-value from the t-test, $p$;  and
the $100(1-\alpha)\%$ confidence interval  on $\beta$ associated with that p-value, $(L,U)$.
\begin{description}
\item[Reproducibility:] Application of the method by two independent statisticians to the same data always gives the same results (as opposed to randomized tests).
\item[Confidence region is an interval:] The confidence region created from $p$ through equation~\ref{eq:CIbyp}
 is an interval,
 meaning it can be written as $(L,U)$.
 \item[Compatible Inferences:]  $p \leq \alpha$ if and only if the $(1-\alpha)$ confidence interval does not contain $\beta_{0}$.
\item[Accuracy (of coverage):] Taken over repeated applications, the probability that the $100(1-\alpha)\%$  confidence interval procedure includes $\beta$ is equal to $(1-\alpha)$ for all values of $\theta$ such that $b(\theta)=\beta$.
  \item[Centrality (of CI):] The $100(1-\alpha)\%$ CI is a central one, meaning $P[L>\beta] \leq \alpha/2$ and $P[U<\beta] \leq \alpha/2$.
 \item[One-sided p-value from Two-sided p-value:] Half of the two-sided p-value can be interpreted as a one-sided p-value in the apparent direction of the effect.
 For example, if $\hat{\beta} > \beta_0$ then we can reject $H_0: \beta \leq \beta_0$ at level $p/2$.
   \item[Directional Coherence (of p-value):]
  Call a two-sided p-value function  {\it directionally coherent} if the p-values are decreasing
  as $\beta_0$ gets farther from  $\hat{\beta}$.
In other words, directionally coherent two-sided p-values have $p({\bf x},\beta_0^*) \leq p({\bf x},\beta_0)$ when either $\beta_0^*< \beta_0 < \hat{\beta}$ or
$\hat{\beta} < \beta_0 < \beta_0^*$.
A two-sided p-value with this property can be interpreted as a coherent one-sided p-value in the appropriate direction.
For example, if $\hat{\beta} > \beta_0$ then we can reject $H_0: \beta \leq \beta_0$
at level $p$. (And for the t-test p-value, we can also reject at a level of $p/2$.)
     \item[Monotonicity (of power):]
     Under the alternative hypothesis, power increases as the sample size increases.
  \item[Nestedness (of CIs):] If  $(1-\alpha^*) >(1-\alpha) $, then the $100(1-\alpha^*)\%$ confidence interval, $(L^*,U^*)$, would  contain the  $100(1-\alpha)\%$ one, $(L,U)$; in  other words, $L^* \leq L < U \leq U^*$.
 \end{description}

\subsection{Two-Sample Binomial: Failure of Normal Intuition}
\label{sec-IntuitionFailure}

Now we turn to the two-sample binomial problem, where
$X_1 \sim Binomial(n_1,\theta_1)$ and independently $X_2 \sim Binomial(n_2,\theta_2)$.
Here the parameter of interest is typically one of three functions of $\theta=[\theta_1,\theta_2]$:
the difference ($\beta_d = \theta_2-\theta_1$), the ratio ($\beta_r=\theta_2/\theta_1$), or the odds ratio
($\beta_{or}=\left\{\theta_2 (1-\theta_1) \right\}/\left\{ \theta_1 (1-\theta_2) \right\}$).
In this problem, the inferential methods do not necessarily follow the properties that we would expect from normal intuition.
We list several examples using several different valid tests, valid confidence intervals, or triples.
\begin{description}
\item[Failure of Reproducibility:] The uniformly most powerful unbiased (UMPU) test of $H_0: \theta_1 \geq \theta_2$ versus $H_1: \theta_1 < \theta_2$ is a randomized version of a one-sided Fisher's exact test
    \citep[see e.g.,][]{Lehm:2005,Finn:2001}.
Testing this hypothesis at the one-sided $\alpha=0.025$ level for the data $x_1/n_1=1/6$ and $x_2/n_2=7/9$, the UMPU test rejects  70.3\% of the time.
So, provided they are not using the same pseudo-random number generator, there is a 41.7\%
chance that two researchers applying the UMPU test to those data will have different accept/reject decisions.
\item[Associated confidence region not an interval:]
There are two versions of the two-sided Fisher's exact test and the most common is the Fisher-Irwin test (default in current versions of SAS [version 9.4]  and R [version 4.0.4], see Section~\ref{sec-twoSidedTests} for definition).
The test was designed to test $H_0: \beta_{or}=1$, but it can be generalized to test other null hypotheses.
Consider the data  $x_1/n_1 = 7/262$ and $x_2/n_2 = 30/494$ \citep[see][Supplement, Section 3.1]{Fay:2010}. The two-sided p-value for testing $H_0: \beta_{or}=1$  is $p=0.04996$, which rejects the null hypothesis at the $\alpha=0.05$ level. If we slightly change the null and test $H_0: \beta_{or}=0.99$, we get $p=0.05005$, and we fail to reject. But  counter-intuitively, if we change the null the other way and test $H_0: \beta_{or}=1.01$, we also fail to reject, $p=0.05006$.
So if we create the $95\%$ confidence region by inverting the p-value procedure, this region is not contiguous,
\begin{eqnarray*}
C({\bf x}, 0.95) & =&  \left\{  \beta: \beta \in (0.177, 0.993) \mbox{ or }
\beta \in (1.006, 1.014) \right\}.
\end{eqnarray*}
and includes values of $\beta_{or}$ both larger and smaller than $1$.
The cause of this behaviour is the lack of unimodality of  the p-value function; see Figure~\ref{figFisherIrwin}.
\begin{figure} 
\vspace{6pc}
\includegraphics[width=5.5in]{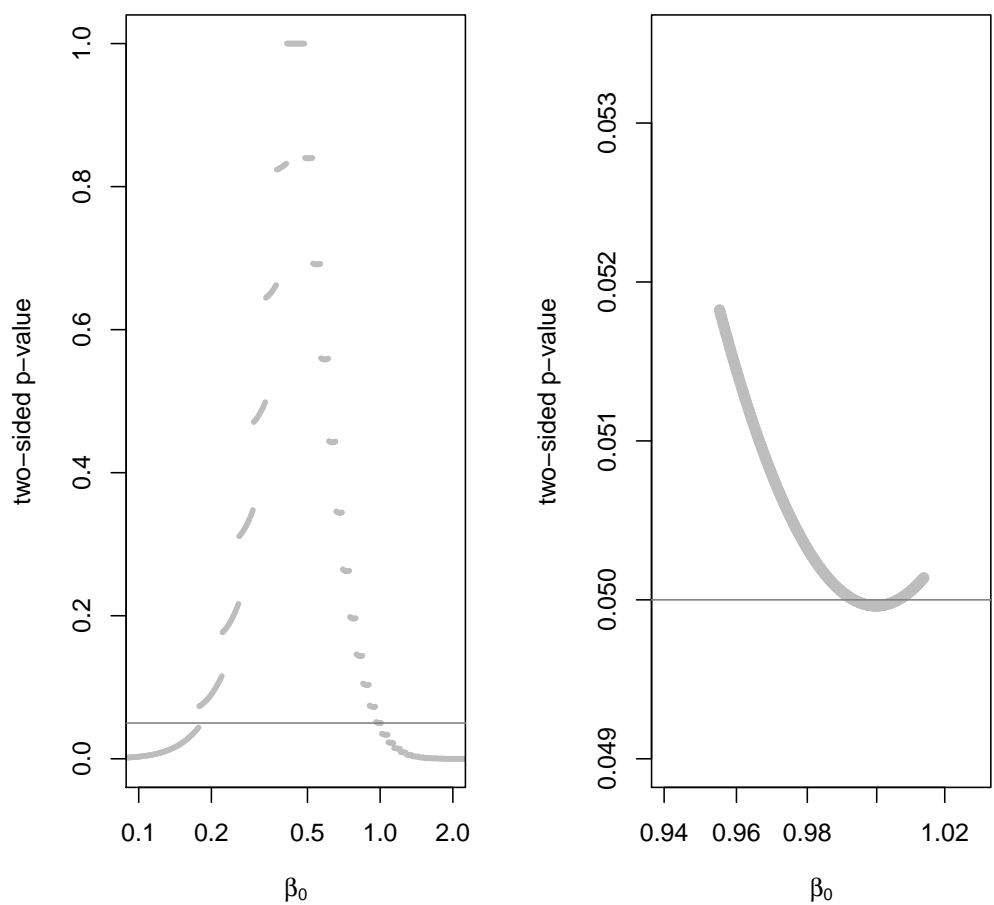}

\caption[]{Two-sided Fisher's exact test (Fisher-Irwin version) p-values by $\beta_0$ for  $x_1/n_1 = 7/262$ and $x_2/n_2 = 30/494$.
Right panel is an enlargement of part of the left panel. Reference line is $0.05$.}
\label{figFisherIrwin}
\end{figure}
\item[Incompatible inferences:] If the confidence region is not an interval, we can create a valid CI by using the interval that covers the whole confidence region.
But this will not give compatible inferences with the p-value function.
Returning to the Fisher's exact test confidence region example,
we can create a 95\% confidence interval by ``filling in the hole''  as $(0.177, 1.014)$ to create the {\it matching} confidence interval \citep[see Section~\ref{sec-definingaMethod} or Ref.][]{Blak:2000}.
In this case, the two-sided p-value rejects the null that $\beta_{or}=1$ at the $0.05$ level, but the matching 95\% confidence interval includes $\beta_{or}=1$.
This issue is different from the incompatible inferences that often occur by using different methods to calculate p-values and confidence intervals, which can be quite prevalent in this application. For example, the default for R ({\sf fisher.test} in base R, version 4.0.4) and SAS (exact option in Proc Freq, version 9.4) uses the Fisher-Irwin two-sided p-value, but calculates the two-sided confidence interval on $\beta_{or}$
by inverting two one-sided Fisher exact p-values
 \citep[see e.g.,][]{Fay:2010,FayR:2010}.
\item[Imperfect Accuracy of Coverage:] Because of discreteness, the valid confidence interval must have coverage larger than the nominal level for some values of $\theta$, in order to ensure validity for all values of $\theta$.
    Remember, the term ``exact'' is often used to mean valid (see Section~\ref{sec-background}), so an ``exact'' confidence interval may have coverage greater than the nominal level and
not, as the term might imply, have coverage exactly equal to the nominal level.
    Section~\ref{sec-midp} discusses relaxing the requirement of validity in order to have coverage closer to the nominal level ``on average'', slightly greater than nominal for some parameter values and slightly less for others.
\item[Non-Centrality of Confidence Interval:] Although central $(1-\alpha)$ CIs for the binomial problem are important, much has been written on non-central intervals. \citet{Agre:2001} showed that inverting certain two-sided tests, produces shorter confidence intervals
    than central ones.
For the difference in proportions, this strategy often uses an unconditional exact (i.e., valid) version of a two-sided score test  \citep[see][]{Fage:2015}.
 For $x_1/n_1=5/9$ and $x_2/n_2=7/7$,  the difference in proportions is $\hat{\beta}_d = 0.444$ with 95\% confidence interval  $(0.005,0.749)$ and the associated two-sided exact p-value for testing $\beta_d=0$ is $p=0.0496$. Because the 95\% confidence interval is based on inverting a two-sided test, we cannot use $p/2=0.0248$ as a one-sided p-value to show that $\beta_d >0$ at the $0.025$ level.
 In fact, to ensure validity, we can only use the two-sided p-value as an upper bound on that one-sided p-value.
\item[Non-monotonicity of power:]
Continuing with the previous example ($x_1/n_1=5/9$ and $x_2/n_2=7/7$ using the unconditional exact two-sided score test), if we add one more observation to group 2 the two-sided p-value increases regardless of whether the extra observation is a  failure (giving $x_2/n_2=7/8$ and $p=0.172$), or success (giving $x_2/n_2=8/8$ and $p=0.0510$)  \citep[this example comes from][]{VosH:2008}.
Thus, it is not surprising that the power to reject at the two-sided $0.05$ level when $\theta_1=.4$ and $\theta_2=.9$ is higher for $n_1=9, n_2=7$ (power= 61.9\%)
than for $n_1=9,n_2=8$ (power=53.7\%).
Power non-monotonicity can also exist for common one-sided tests.
Using a one-sided Fisher's exact test (to reject $H_0: \theta_1 \leq \theta_2$) at the $0.025$ level, when $n_1=n_2=4$ the only
way to reject is the most extreme case, $x_1/n_1=4/4$ and $x_2/n_2=0/4$, and the power to reject is $(1-\theta_1)^4 \theta_2^4$.  When $n_1=4$ and $n_2=5$ then still the only way to reject at the $0.025$ level is the most extreme case, $x_1/n_1=4/4$ and $x_2/n_2=0/5$, and the power is $(1-\theta_1)^4 \theta_2^5$.
Then, $(1-\theta_1)^4 \theta_2^4> (1-\theta_1)^4 \theta_2^5$ for all $0<\theta_1<1$ and $0<\theta_2<1$, and the power with the larger sample size is smaller. A similar decrease in power occurs by instead adding one to the other group: $n_1=5$ and $n_2=4$.
\item[Non-nested Confidence Intervals:] \citet{Wang:2010} proposed a method for constructing the smallest one-sided confidence interval for the difference of two proportions.
Consider $x_1/n_1=2/7$ and $x_2=2/5$. The lower one-sided 95\% interval on the difference, $\beta_d$, is $(-0.467,1)$, but the 96\% interval by the same method is
$(-0.442,1)$.  See Figure~\ref{figWangDiff} and Section~\ref{sec-Talpha}.
\begin{figure} 
\vspace{6pc}
\includegraphics[width=5.5in]{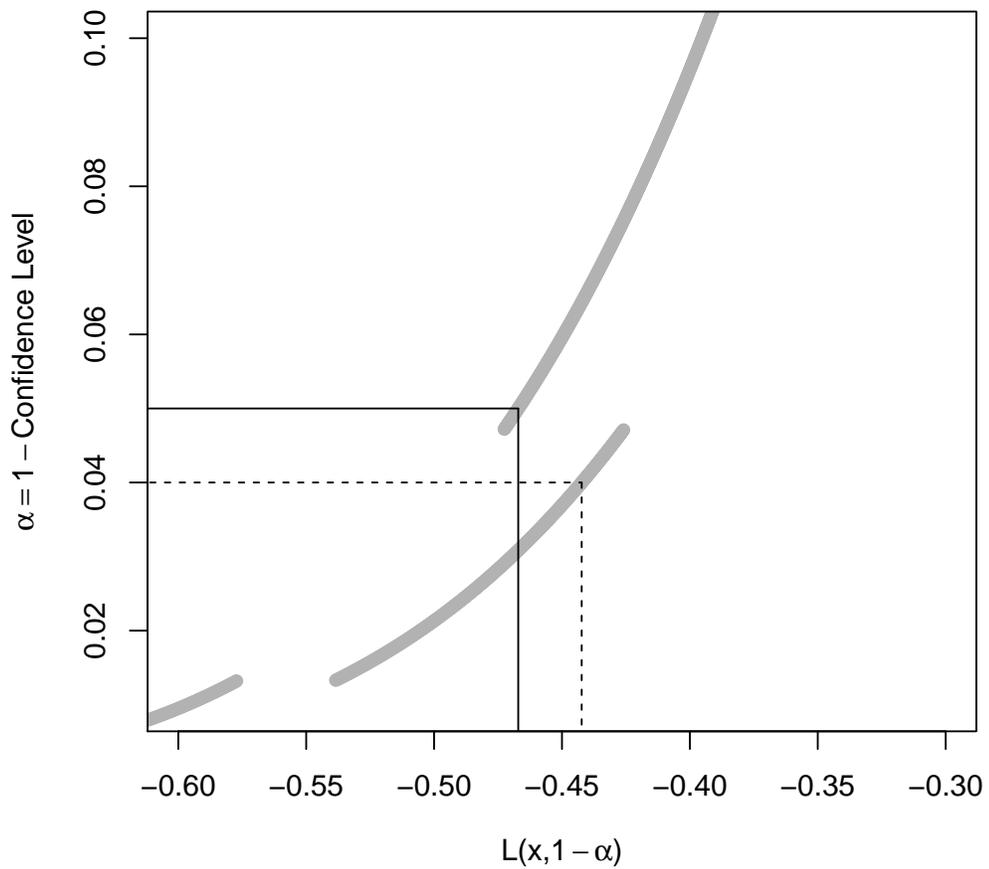}

\caption[]{Thick gray lines are lower limits for the smallest one-sided  $100(1-\alpha)\%$ confidence limits for $\beta_d$ from \citet{Wang:2010} for $x_1/n_1=2/7$ and $x_2/n_2=2/5$.
Solid black lines show one-sided 95\% limit of $-0.467$, while dotted black lines show one-sided 96\% limit of $-0.442$.
}
\label{figWangDiff}
\end{figure}
\item[Non-Coherence:]
For testing for non-inferiority on a difference in proportions, \citet{Chan:1999} recommend the exact unconditional test based on the score test.
\citet{Rohm:2005} gives the following illustrative example: the proportion of failures on control is $x_1/n_1= 130/248$ and on new treatment is $x_2/n_2=76/170$, with the failure rate slightly lower on new treatment, $\hat{\beta}_d =-0.077$. If we want to show  that $H_1: \beta_d < 0.025$ the p-value is $p=0.0226$, but if we want to show an even less stringent margin, $H_1: \beta_d < 0.026$ the p-value non-intuitively  increases to $p=0.0240$ (see Figure~\ref{figRohmelExample} and Section~\ref{sec-Tbeta}).
\begin{figure} 
\vspace{6pc}
\includegraphics[width=5.5in]{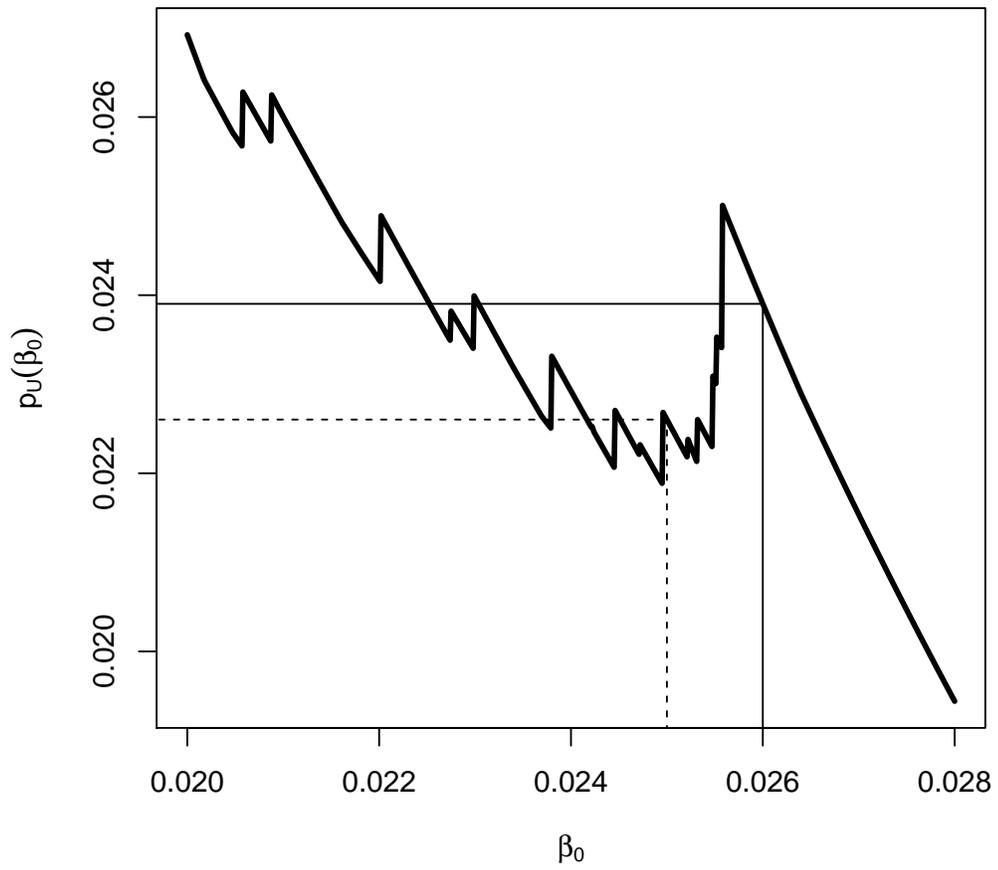}

\caption[]{One-sided exact unconditional p-value using the score statistic ordering, $p_U({\bf x}, \beta_0)$. Virtual data example from \citet{Rohm:2005}:
$x_1/n_1=130/248$ and $x_2/n_2=76/170$. Thick black line is the p-value function. Thin black lines depict the test of the null $H_0: \beta \geq 0.026$
and dotted lines depict the test of the null $H_0: \beta \geq 0.025$.  }
\label{figRohmelExample}
\end{figure}
\end{description}

For the two-sample binomial problem, many attempts to increase power or get the smallest expected length CI
result in violations of some of these  ``normal intuition'' properties.

\section{Choosing the Effect Measure}
\label{sec-EffectChoice}

Choosing the effect measure is dependent on the application, so we examine a real application to discuss the issues.
\citet{Coul:2009} studied a parasite called {\it Mansonella perstans} that infects people in parts of Africa.  The usual drugs that kill other similar parasites had not been working on killing {\it M. perstans}.  \citet{Coul:2009} realized that in this case there was a symbiotic bacteria, {\it Wolbachia}, that helped the {\it M. perstans} live. They suspected that if they gave a common antibiotic, doxycycline, to kill the bacteria, it may in fact help cure the patient of {\it M. perstans}.  Patients were randomized to the treatment group (doxycycline) or the control group (no treatment). There are issues of missing data that we ignore for simplicity. At 12 months, $x_2=67$ out of $n_2=69$ subjects who received doxycycline and $x_1=10$ out of $n_1=63$ had cleared the {\it M. perstans} from their blood.
There are several reasonable choices for how to measure the effect: the difference in clearance rates, the ratio of clearance rates, the ratio of failure probabilities,
and the odds ratio of clearance rates. Although the choice is often dominated by what is most natural to the intended audience, there are some statistical issues related to this choice.


Without loss of generality, consider effect measures that measure how much larger $\theta_2$ is than $\theta_1$. The opposite effect can be measured by switching group labels.
But we could also simultaneously switch group labels {\it and } switch the response and failure labels. If the effect remains the same after this double switching, we say that the measure has symmetry equivariance. The measures $\beta_d$ and $\beta_{or}$ have symmetry equivariance; however, $\beta_r$ does not have it,
as we demonstrate with the example.
Let  $\hat{\theta}_2 = 67/69 \approx 0.97$ and $\hat{\theta}_1=10/63 \approx 0.16$. An estimate of the rate ratio for success (cleared parasites at 12 months) is $\hat{\theta}_2/\hat{\theta}_1 \approx 6.12$. The rate ratio is often called the relative risk, but in this case the ``risk'' is the risk of getting cured. A different expression of the same data measures the ratio of the rates of failures (those still having detectable parasites at 12 months). Let $\hat{\theta}_{F2} = 2/69 \approx 0.03$ and $\hat{\theta}_{F1} = 53/63 \approx 0.84$. Then an estimate of the relative risk of failure   is $\hat{\theta}_{F1}/\hat{\theta}_{F2} \approx 29.0$. In this latter case the control group looks about 29 times worse than the treatment group, while if we look at the rate ratios for success, the treatment group looks only about 6 times better than the control group. So how many times better treatment is than control depends on which way we measure risk.
This is a violation of symmetry equivariance.
Despite this, the rate ratio is often used because it is easy to understand \citep[see e.g.,][]{Coul:2009}, or because it has become the parameter of choice within a field, so that its use facilitates comparisons between studies.

The difference has symmetry equivariance.  If we measured the difference in rates of disease rather than the difference in  rates of cure we get exactly the negative difference as we might expect.  Similar to the relative risk, the difference is often used because it is easy to understand. Additionally, the sample difference in rates is always defined, unlike the ratio which is undefined when $\hat{\theta}_1=\hat{\theta}_2=0$.

\begin{figure} 
\vspace{6pc}
\includegraphics[width=5.5in]{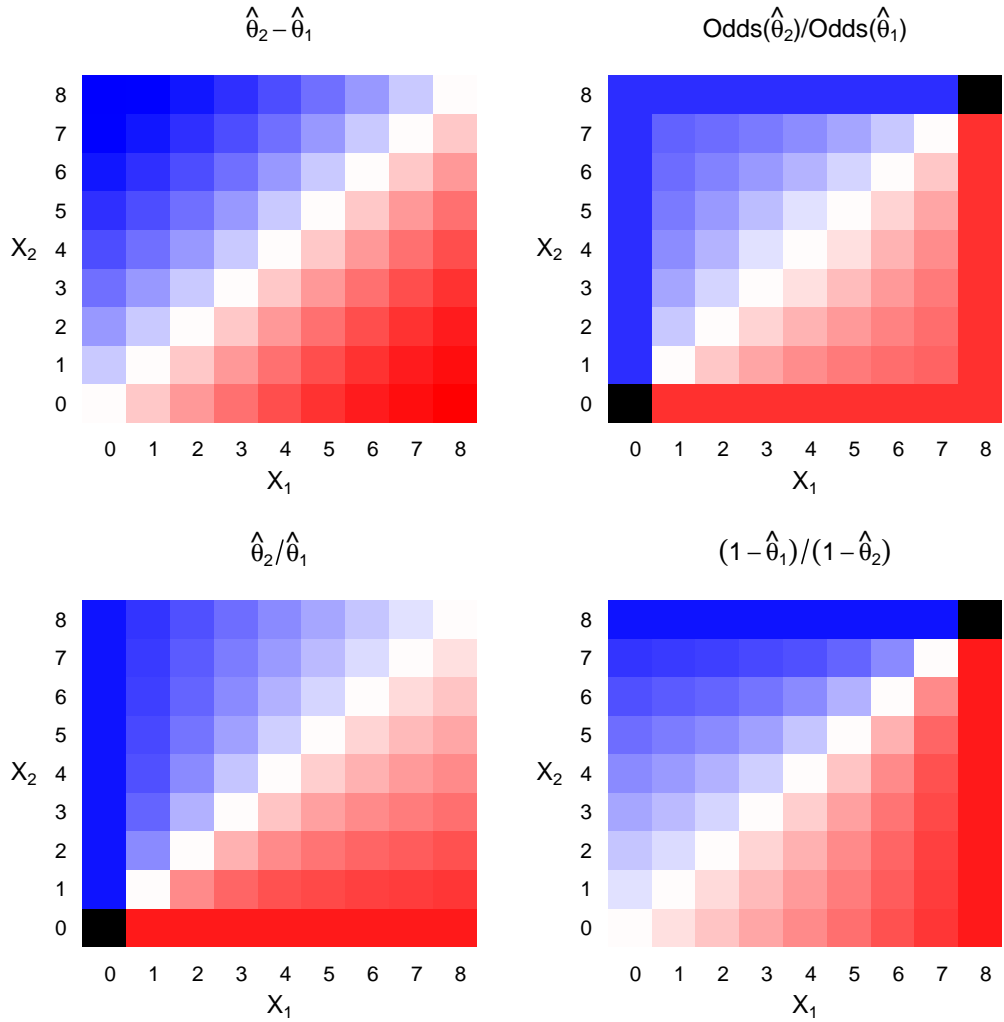}

\caption[]{Four simple ordering functions. Dark blue means $\hat{\theta}_2$ is much larger than $\hat{\theta}_1$ and dark red is the opposite. White means both treatments appear the same.
The functions are based on $n_1=n_2=8$ and using functions of the sample proportions, $\hat{\theta}_1=x_1/n_1$  and $\hat{\theta}_2=x_2/n_2$.
The sample space is depicted by a $9 \times 9$ grid of responses,
 ranked by the ordering functions: difference in success proportions (upper left), odds ratio (upper right), ratio of success proportions (lower left), and
 ratio of failure proportions (lower right). Colors rank the functions from the highest values (dark blue) indicating larger $\theta_2$, to middle values (white) indicating
 $\theta_1=\theta_2$, to lowest values (dark red), with black indicating no information.
}
\label{figbetaPlots8}
\end{figure}

Figure~\ref{figbetaPlots8} plots the three statistics using $\hat{\theta}_2$ and $\hat{\theta}_1$ with $n_1=n_2=8$.
The plots go from dark blue ($\hat{\theta}_2$ is larger) to white ($\hat{\theta}_1=\hat{\theta}_2$) to dark red ($\hat{\theta}_1$ is larger), with black denoting indeterminate.
Because of the indeterminate black areas, the ordering of the sample space for the ratio and odds ratio is not straightforward (see Section~\ref{sec-orderingRatio}).
The ordering of the measures on the parameters themselves would give a continuous version of Figure~\ref{figbetaPlots8}, and the black regions would reduce to points at $(\theta_1,\theta_2)=(0,0)$ or $(1,1)$.
The bottom panels show the lack of symmetry equivariance for $\beta_r$. Comparing the panel for $\beta_{or}$ with the two different ratio panels, we see that the lower left hand corner of the $\beta_{or}$ panel is similar to the lower left hand corner of $\hat{\beta}_r = \hat{\theta}_2/\hat{\theta}_1$. For small $\theta$, $\hat{\beta}_{or}$ is a good approximation to $\hat{\beta}_r$. Similarly for both $\theta$ values close to $1$, $\hat{\beta}_{or}$ is a good approximation of $(1-\theta_1)/(1-\theta_2)$ (right bottom panel).

The odds ratio is the most complicated of the three measures, but it has some nice properties. It is very important for the case-control design used to study rare diseases,
because the odds ratio of disease given exposure is equal to the odds ratio of exposure given disease \citep[see][]{Bres:1996}.
Also for performing regression on binary observations, logistic regression allows linear predictors to be used to model the log odds,
and effects of binary covariates can be expressed
as odds ratios.  An advantage of the odds ratio for the two-sample binomial case is that by conditioning on the total number of successes in both groups, the probability distribution
reduces to a noncentral hypergeometric distribution which is a function of $\beta_{or}$. This is discussed more in Section~\ref{sec-oneSidedCond}.

\section{Properties of Frequentist Triples}
\label{sec-methods}

\subsection{Defining a Matched Triple}
\label{sec-definingaMethod}

Once we choose an effect measure, we choose an appropriate {\it triple} (an estimator, confidence interval, and p-value function) for inferences.
We will not specify the estimator except to require that it is within the confidence interval.
We focus mostly on choosing the CI and p-value function.
Except in Section~\ref{sec-midp}, we only consider  triples that are valid (i.e., the CI and p-value are both valid) and reproducible.
Because we require reproducibility, the triple based on the UMP unbiased (and randomized) test  is not allowed.
We focus on triples where the p-value function and the confidence interval are derived from the same procedure. We call this a {\it matched triple}.

Here is a precise definition of a matched triple.
 If we start with $p({\bf x}, \beta_0)$, an associated confidence region is given by equation~\ref{eq:CIbyp},
 and  the matching CI is smallest interval that
contains that  confidence region.  In other words, if the confidence region has holes in it, then those holes are ``filled in''.
On the other hand, if we start with $(L,U)= C_{I}({\bf x}, 1-\alpha)$, then  the matching p-value function is
the smallest $\alpha$ such that $\beta_0$ is outside $C_{I}({\bf x}, 1-a)$ for all $a \geq \alpha$.

\subsection{Implications of Compatible Inferences}

\begin{thm}
\label{THM}
Consider a  valid, reproducible, and matched triple. The triple has compatible inferences
\begin{enumerate}
\item
\label{thmUCR}
if and only if the CI is equal to the confidence region associated with the p-value, and
\item
\label{thmUN}
only if the CI is nested, and
\item
\label{thmUOS}
only if the the p-value function is coherent (for one-sided p-values),
or directionally coherent (for two-sided p-values).
\end{enumerate}
\end{thm}

The formal proof of the theorem is in Appendix~\ref{sec-proof}.
The theorem says we must have nested CIs and coherent p-values in order to have compatible inferences.
These ideas are best understood graphically.
Figure~\ref{figFisherIrwin}
shows lack of directional coherence; for every $\beta_0$ there is only one p-value, and the two-sided p-value function is not unimodal (i.e., as $\beta_0$ increases, the p-value function does not increase to the global maximum, then decrease after that; see the right panel).
Similarly, Figure~\ref{figRohmelExample} shows lack of coherence.
Figure~\ref{figWangDiff} shows non-nestedness;  for every $\alpha$ there is only one lower limit, and the lower limit is not a
monotonic function of the level.


\subsection{Directional Inferences}
\label{sec-3sided}

Typically, a researcher who finds a significant difference from the two-sided p-value suggesting that $\beta \neq \beta_0$ is almost always interested in
interpreting the result in terms of whether $\beta>\beta_0$ or $\beta<\beta_0$. In other words, the two-sided hypothesis test is often treated as a three-decision rule:
(1) fail to reject $\beta=\beta_0$, (2) reject $\beta=\beta_0$ and conclude $\beta>\beta_0$, or (3)  reject $\beta=\beta_0$ and conclude $\beta<\beta_0$.
If the two-sided p-value has directional coherence, then if we reject $H_0: \beta=\beta_0$ at level $\alpha$, we can additionally reject at level $\alpha$ either
$H_0: \beta \leq \beta_0$ (if $\beta_0 < \hat{\beta}$) or   $H_0: \beta \geq \beta_0$ (if $\beta_0 > \hat{\beta}$).

Consider comparing two triples that both have compatible inferences, one with a central CI, and one with a non-central CI.
For the non-central triple (i.e., the one with the non-central CI) the associated two-sided hypothesis test may be slightly more powerful, but if the non-central triple is applied also to a subsequent one-sided hypothesis (as in the three decision rule), it can be quite a bit less powerful than the central one.
To see this,
start with a nested central CI, say $(L,U)$, and pair it with its matching two-sided p-value, say $p_C$.
By Theorem~\ref{THM}, this means that
whenever the $100(1-\alpha)\%$ CI  excludes $\beta_0$ then $p_C \leq \alpha$, and we can reject $H_0: \beta=\beta_0$ at level $\alpha$.
After rejecting the two-sided hypothesis at level $\alpha$, we can reject one of the one-sided hypotheses at level $\alpha/2$;
if $\beta_0 < L$ we reject $H_0: \beta \leq \beta_0$, while if  $\beta_0 > U$ we reject $H_0: \beta \geq \beta_0$. A non-central
CI does not allow one-sided rejections at the $\alpha/2$ level.
\citet{Free:2008} discusses this issue in terms of clinical trials, and, using these arguments as well as some Bayesian motivation, \cite{Free:2008} recommends performing two one-sided tests at the $\alpha/2$ level, which is another way of describing the use of central CI methods for three decision rules.

In summary, if we desire directional inferences, and we want to compare the power to detect a one-sided effect in a fair way (i.e., both methods bound the one-sided type I error rates of the three decision rule at the same level),
then we need to compare a method with a two-sided p-value and its matching $100(1-\alpha)\%$ non-central CI,
with  a pair of one-sided p-values and its matching $100(1-2 \alpha)\%$ central CI. This means that when comparing expected lengths of CIs,
if directionality of effect is important, we should compare the expected length of a $100(1-\alpha)\%$ non-central CI with the expected length of a
$100(1-2\alpha)\%$ central CI. Because directionality is usually important, our default recommendation is to use central confidence intervals
and perform three-sided inferences as described above.

\section{Methods for Creating One-Sided Exact Unconditional Testing Procedures}
\label{sec-OnesidedUncond}


\subsection{Basic Procedure for Defining p-values}
\label{sec-UncondBasic}

Suppose larger values of $\theta$ are better. We want to know if treatment 2 is better than treatment 1 ($\theta_2>\theta_1$), and by how much.
Let $T({\bf x})$ be a function of the data, where larger values of $T({\bf x})$ indicate that treatment 2 is better than treatment 1,
and $T({\bf X})$ is defined for all possible values of ${\bf X}$.
For example, a simple $T({\bf x})$ is the difference in observed proportions (see Figure~\ref{figbetaPlots8} upper left).
For this section and the next (Section~\ref{sec-refinement}), we require that $T$ is a function of ${\bf x}$ only.
Later in Section~\ref{sec-Talpha}, $\;T$ may depend on $\alpha$, and in Section~\ref{sec-Tbeta}, $\;T$ may depend on $\beta_0$.
\citet{Barn:1947} outlined convexity conditions which ensure
that larger values of $T$ suggest treatment 2 is better.
Barnard's convexity (BC) conditions are:
\begin{eqnarray}
  \mbox{ if $x_2^*>x_2$} &  \mbox{then}  &  T([x_1,x_2^*]) \geq T([x_1,x_2])  \nonumber \\
& \mbox{and} &   \label{eq:Barnard} \\
      \mbox{ if $x_1^*<x_1$} & \mbox{then} & T([x_1^*,x_2]) \geq T([x_1,x_2]).   \nonumber
\end{eqnarray}
Many choices for $T$ satisfy the BC conditions.
For example,
$T({\bf x}) = \hat{\theta}_2-\hat{\theta}_1$
 meets the BC conditions.

Once we have decided on the ordering function, $T$, we can create valid unconditional one-sided p-values: $p_U$  for testing the null $H_{U0}$
(defined as $H_0: \beta \geq \beta_0$)
and $p_L$ for testing $H_{L0}$  ($H_0: \beta \leq \beta_0$) using
\begin{eqnarray}
p_{U}({\bf x}, \beta_0) & = & \sup_{
\theta: b(\theta) \geq \beta_0
} P_{\theta} \left[ T({\bf X}) \leq T({\bf x}) \right] \nonumber \\
& \mbox{and} & \label{eq:Tpvalues} \\
p_{L}({\bf x}, \beta_0) & = & \sup_{  \theta: b(\theta) \leq \beta_0
} P_{\theta} \left[ T({\bf X}) \geq T({\bf x}) \right]. \nonumber
\end{eqnarray}
These p-values are valid since
\begin{eqnarray*}
\sup_{\theta \in \Theta_0} P_{\theta} [ p({\bf X},\beta_0) \leq p({\bf x},\beta_0) ] \leq p({\bf x},\beta_0)
\end{eqnarray*}
where $\Theta_0 = \left\{ \theta: b(\theta) \geq \beta_0 \right\}$ for $p_U$ and
$\Theta_0 = \left\{ \theta: b(\theta) \leq \beta_0 \right\}$ for $p_L$.
The p-values are also `exact' by the terminology of \cite{Ripa:2017} (see  equation~\ref{eq:RipaExact}, or Theorem 1 of \cite{Lloy:2008}).
Thus, any other valid p-values
with ordering based on $T$
are inadmissible (that is, they have values that are never less than the valid unconditional p-values and are greater  for at least one ${\bf x}$) \citep[][Theorem 2]{Lloy:2008}.

These valid one-sided p-values can be inverted to create two $100(1-\alpha/2)$ one-sided confidence limits using
\begin{eqnarray}
U({\bf x})
& = &
\left\{
\begin{array}{ll}
\sup  \left\{ \beta_0:  \right. & \left. p_{U}({\bf x}, \beta_0) > \alpha/2  \right\},
\mbox{ if $\exists$ a $\beta_0$  }    \\
&
\mbox{with $p_{U}({\bf x}, \beta_0) > \alpha/2$}  \\
\beta_{max} & \mbox{otherwise}
\end{array}
\right.
\nonumber \\
& &  \hspace*{5em} \mbox{and} \label{eq:onesidedCIs} \\
L({\bf x})
& = &
\left\{
\begin{array}{ll}
\inf  \left\{ \beta_0: \right. & \left. p_{L}({\bf x}, \beta_0) > \alpha/2  \right\},
\mbox{if $\exists$ a $\beta_0$}
  \\
& \mbox{  with $p_{L}({\bf x}, \beta_0) > \alpha/2$}  \\
\beta_{min} & \mbox{otherwise}
\end{array}
\right.
\nonumber
\end{eqnarray}
where $(\beta_{min},\beta_{max}) = (-1,1)$ for $\beta_d$ and $(0,\infty)$ for $\beta_r$ or $\beta_{or}$.
A central $100(1-\alpha)$ confidence interval is the union of the one-sided ones,  $( L({\bf x}), U({\bf x}))$,
and a central p-value is $p_C({\bf x},\beta_0) = \min(1,2 p_L, 2 p_U)$.
These confidence limits are called exact unconditional \citep[see e.g.,][]{Mehr:2003} or Buehler confidence limits
\citep[see][]{Lloy:2003}.
\citet{Lloy:2003} and \citet{Wang:2010}
 show two results about these one-sided intervals.
First,
the lower and upper one-sided confidence limits retain a logical ordering analogous to Barnard's convexity conditions.
Specifically, $(L,U) \in \mathcal{O}_T$, where $\mathcal{O}_T$ is the class of valid central
confidence intervals such that
 if $T({\bf x}_1) < T({\bf x}_2)$ then $L({\bf x}_1) \leq L({\bf x}_2)$ and
$U({\bf x}_1) \leq U({\bf x}_2)$.
Second, $(L, U)$ calculated in this manner is the smallest confidence interval within $\mathcal{O}_T$.
In other words, any other valid central
confidence interval $(L^*, U^*)$  in $\mathcal{O}_T$  must have  $L^*({\bf x}) \leq  L({\bf x})$ and $U({\bf x}) \leq  U^*({\bf x})$
for all ${\bf x} \in \mathcal{X}$.

\citet{Barn:1947} proposed the two-sided CSM test (we discuss the name later). We define the CSM test more generally using an ordering method which may be used for one or two-sided tests
and confidence intervals, and we begin with the one-sided versions.
Briefly, Barndard's CSM one-sided ordering starts from the most extreme point and incrementally adds more points to the order such that (1)  the new point(s) and all previous points meet the BC conditions, and (2)
the new point(s) have the lowest one-sided p-value among the possible new points that meet the BC conditions.
Details are in Appendix~\ref{sec-BarnardCSM}. Once the appropriate one-sided CSM ordering function, $T({\bf x})$, is defined, we use the above definitions for the p-values (equation~\ref{eq:Tpvalues})
and confidence intervals (equation~\ref{eq:onesidedCIs}).
The CSM stands for convexity, symmetry, and maximum. {\it Convexity} refers to the BC condition that each new point must meet, and {\it maximum} refers to the maximization of the null hypothesis space
in the definition of the p-value (see sup expression in equations (\ref{eq:Tpvalues})).  The {\it symmetry} condition only applies to the two-sided version of the CSM test, but nevertheless we use ``CSM'' to describe all versions. The symmetry condition states that whenever
a point $[x_1^*,x_2^*]$ is added to the order, one must simultaneously add $[n_1-x_1^*, n_2-x_2^*]$ and give it the same $T$ value in the ordering (see Appendix~\ref{sec-BarnardCSM} and Section~\ref{sec-twoSidedTests} for more discussion of two-sided tests).

In a different paper, \citet{Barn:1945} outlined the general exact unconditional test, and those tests are sometimes referred to as ``Barnard's test'' \citep[see e.g.,][]{SAS:2012,Cyte:2010},
but we do not use that terminology to avoid confusion with Barnard's CSM test.
\citet{Rohm:2013} discussed one-sided exact unconditional tests using Barnard's CSM p-value ordering, except with  breaking more ties to get higher power,
an idea discussed in the next section.

\citet{Mart:1998} proposed a good all-purpose ordering, which is to base the ordering on the one-sided mid-pvalue from Fisher's exact test (see equation~\ref{eq:pcumid}). We explore the power properties of this ordering in Section~\ref{sec-power}. Alternatively, the ordering can be tailored to a specific application. For example, \citet{Gabr:2018} proposed an ordering to optimize power for certain types of animal experiments where $\theta_1$, the parameter for the control group, is expected to be nearly 1.

\subsection{Improving Power by Breaking Ties: Refinement of Ordering Functions}
\label{sec-refinement}

One important way to improve the power of some unconditional exact tests based on a function $T$
is to break any ties that exist in the ordering function.
If $T$ is an ordering function with ties, and $T^*$ is an ordering function that gives the same ordering of $T$ at all the untied values and additionally breaks some ties,
then we say $T^*$ is a {\it refinement} of $T$.
Then the unconditional exact p-values formed with $T^*$ are always less than or equal to those formed with $T$ \citep[see][p. 158]{Rohm:1999}.
Similarly, one-sided exact unconditional lower confidence limits formed using $T^*$ are always at least as large as the ones formed using $T$ \citep{Kaba:2006,Wang:2010}.

We describe one specific refinement or tie breaking algorithm for  the difference in proportions next, which
as far as we are aware, has not been specifically described in the literature and has not been available in software
(although there are some closely related methods).
We can order within each set of tied values using Wald statistics for $\hat{\beta}_d$, i.e., ordering by
\begin{eqnarray*}
Z({\bf x}) & = & \frac{\hat{\beta}_d}{\sqrt{\widehat{var}_0(\hat{\beta}_d)}} =  \frac{  \hat{\theta}_2 - \hat{\theta}_1   }{ \sqrt{ \hat{\theta}(1-\hat{\theta})( 1/n_1+ 1/n_2 )} }
\end{eqnarray*}
where $\hat{\theta}=(x_1+x_2)/(n_1+n_2)$.
This leaves the ties for $\hat{\beta}_d=0$, but otherwise defines points with more precision as more extreme, where extreme is further away from zero.
Not all the values with $\hat{\beta}_d \neq 0$ break all the ties.
For example, consider the ties at  $\hat{\beta}_d=5/8$ that happen at the ${\bf x}$ values  $[0,5]$, $[1,6]$, $[2,7]$, and $[3,8]$, for $n_1=n_2=8$.
This method still leaves tied the two pairs of points,  $\left\{[0,5], [3,8]\right\}$ and $\left\{ [1,6], [2,7] \right\}$. These remaining ties we argue
 should remain tied in order for the ordering to retain symmetry equivariance.
Note that this suggested ordering is similar, but not equivalent to just ordering the entire sample space by $Z({\bf x})$ \citep[as was studied in][]{Mehr:2003}.

If we break the ties in this way, then the BC conditions are still met,
because only at the boundaries (where the ties are broken according to the BC conditions)
do the ties occur at two points ${\bf x}_a$ and ${\bf x}_b$ with $x_{a1}=x_{b1}$ or $x_{a2}=x_{b2}$.
All of the other ties will not have any  $x_{a1}=x_{b1}$ or $x_{a2}=x_{b2}$ so they can be broken in any manner and the overall ordering function, $T^*$,
will meet the BC conditions. This is important for computation (see Section~\ref{sec-Computations}).
Further, the proposed $T^*$ (tie-breaking on difference in proportions) does not depend on $\alpha$ or $\beta_0$ like some score test based methods
(see Sections~\ref{sec-Talpha} and \ref{sec-Tbeta}) so avoids problems with nesting and coherence.


\subsection{Ordering Functions for Ratio and Odds Ratio}
\label{sec-orderingRatio}

Performing exact unconditional tests on $\beta_r$ or $\beta_{or}$ is not straightforward. We consider $\beta_r$ first since it is simpler.
One problem is that  ${\bf x} = [0,0]$ could occur with high probability if the true ratio was $100$ or if it was $1/100$ as long as both $\theta_1$ and $\theta_2$ were very small. So if $T({\bf x})$ is designed so that larger values suggest $\theta_2 > \theta_1$, it is not clear  how to define $T([0,0])$ if our interest
is in $\beta_r$.

Since ${\bf x} = [0,0]$ gives no information  about $\beta_r$, we must deal with $[0,0]$ in a special way; set the p-value at ${\bf x}=[0,0]$ to $1$ for tests of $\beta_r$ regardless of the null hypothesis. This means that ${\bf x}=[0,0]$ is placed ``deepest'' within the null.
Following equations~\ref{eq:Tpvalues}, this implies $T([0,0])$ can be thought of as the largest value when calculating $p_U({\bf x}, \beta_0)$
and the smallest value when calculating   $p_L({\bf x}, \beta_0)$. A similar issue applies to the odds ratio, except in that case, the point ${\bf x}=[n_1,n_2]$ also has no  information about $\beta_{or}$.

For clarity, we rewrite  equations~\ref{eq:Tpvalues} applied to all three parameters. Let $\mathcal{X}_I$ denote the set of ${\bf X}$ values with information about $\beta$.
Then if  ${\bf x} \notin \mathcal{X}_I$ set $p_U({\bf x}, \beta_0)$ and $p_{L}({\bf x}, \beta_0)$ to $1$, otherwise
let $p_U({\bf x}, \beta_0)$ be
\begin{eqnarray}
 \sup_{
\theta: b(\theta) \geq \beta_0
} P_{\theta} \left[ T({\bf X}) \leq T({\bf x}) | {\bf X} \in \mathcal{X}_I \right]  P_{\theta} \left[ {\bf X} \in \mathcal{X}_I  \right] \nonumber
\end{eqnarray}
and analogously, let  $p_L({\bf x}, \beta_0)$ be
\begin{eqnarray}
 \sup_{
\theta: b(\theta) \leq \beta_0
} P_{\theta} \left[ T({\bf X}) \geq T({\bf x}) | {\bf X} \in \mathcal{X}_I \right]  P_{\theta} \left[ {\bf X} \in \mathcal{X}_I  \right]. \nonumber
\end{eqnarray}
Since we never reject when ${\bf x} \notin \mathcal{X}_I$, these definitions give valid p-values, and additionally when ${\bf x} \notin \mathcal{X}_I$ we do not need to define $T({\bf x})$.

The simple ordering function by the estimate of $\beta_r$ or $\beta_{or}$ (even when using
a tie breaking ordering similar to what was done for $\beta_d$)  is not very powerful (see Section~\ref{sec-power}), and is not recommended. Typically,
we order using a score function (see Section~\ref{sec-Tbeta}) since it gives more reasonable power.

\subsection{Other Improvements: E+M and Berger-Boos}
\label{sec-otherImprovements}

Another method to apparently improve the ordering statistic for any efficacy parameter (difference, ratio, or odds ratio) is the estimated and maximized ($E+M$) p-value \citep{Lloy:2008}.
In this method, we replace an ordering statistic, $T$,
with $T^*$, where $T^*$ is an estimated p-value when testing $H_{L0}$ (or the negative estimated p-value when testing $H_{U0}$).
We estimate the p-value by plugging in
$\hat{\theta}_0$ instead of taking the supremum of $\theta$ under the null, where
 $\hat{\theta}_0$ is the  maximum likelihood estimator of
$\theta \in \Theta_0$.
For example, the approximation for $p_L$ in expression~\ref{eq:Tpvalues} uses $\hat{p}_L({\bf x}, \beta_0) = P_{\hat{\theta}_0} \left[ T({\bf X}) \leq T({\bf x}) \right]$. Then we ``maximize'' using $T^*({\bf x}) = \hat{p}_L({\bf x}, \beta_0)$
instead of $T$ as the ordering function.  That is, we calculate the exact conditional p-value using expression~\ref{eq:Tpvalues} by taking the supremum.
\citet{Lloy:2008} studied this method and observed that when $T^*$ (the approximate p-value) is used as the ordering statistic,  the
resulting exact unconditional p-value is generally smaller than the exact unconditional p-value on $T$.
The process can be repeated (replace $T^*$ by its approximate p-value), but the additional reduction appears to be minimal.

\citet{Berg:1994} introduced a popular adjustment that tends to reduce exact unconditional p-values. Instead of taking the supremum over the entire null hypotheses parameter space,
 take the supremum only over $C_{\gamma}$, a $100(1-\gamma)\%$ confidence set of $\theta$ restricted to be in the null space, then add $\gamma$ to ensure validity.
 This is usually done by reexpressing the parameter space $(\theta_1,\theta_2)$ as $(\beta, \psi)$, where $\psi$ is a nuisance parameter, then defining $C_{\gamma}$ as the intersection of  $\theta \in \Theta_0$ and the set of $\theta$ values with $\psi$ in its $100(1-\gamma)\%$ confidence interval.
A Berger-Boos version of $p_U$ of expression~\ref{eq:Tpvalues}, uses
\begin{eqnarray*}
p_{U\gamma}({\bf x}, \beta_0)  & = & \gamma + \sup_{\theta \in C_{\gamma}}  P_{\theta} \left[ T({\bf X}) \geq T({\bf x}) \right].
\end{eqnarray*}
This is not optimal, since we may be able to improve it by using $p_{U\gamma}({\bf x}, \beta_0)$ as an ordering function. Nevertheless, it usually provides some reduction in p-values
\citep[see e.g.,][]{Lloy:2008}.

\subsection{Ordering Functions That Depend on Significance Level}
\label{sec-Talpha}

\citet{Kaba:2003} showed that for one-sided $100(1-\alpha/2)\%$  exact unconditional upper confidence limits, the ordering function, $T$, that maximizes the asymptotic efficiency is an approximate $100(1-\alpha/2)\%$ one-sided upper confidence limit itself. A different ordering function is used for the upper and lower limit, and for different confidence levels.

\citet{Wang:2010} and \citet{Wang:2015} also proposed an ordering function to give the smallest CI, and the calculation of the ordering function itself is
iterative and
 quite involved, similar to the CSM test of \citet{Barn:1947}.
The precise definition of the ordering is notationally cumbersome, but the idea is roughly as follows.
Consider the lower $100(1-\alpha/2)\%$ one-sided limit.
Start from the most extreme point ${\bf x} = [0,n_2]$. Then
add points one at a time, picking the point, ${\bf x}_a$,  that gives  the largest $L({\bf x}_a, 1-\alpha/2)$ and belongs to the set of closest neighboring points with the already included points, where closest neighbor is defined in terms of the BC conditions. The algorithm ensures that the lower limit function meets the BC conditions. Because each added $L({\bf x})$ value is as large as possible,  if the resulting ordering function $T$ gives the finest partition (there are no ties), then any valid $100(1-\alpha/2)\%$ one-sided lower limit that meets the BC conditions and uses $T$ for ordering, say $L^*$, has $L^*({\bf x}) \leq L({\bf x})$ for all ${\bf x}$ \citep[see][]{Wang:2010,Wang:2015}.

The price for this  optimality property is that the ordering function depends on $\alpha$. Different ordering functions arise for different $\alpha$, which can lead to non-nestedness (see Figure~\ref{figWangDiff}).

\subsection{Ordering Functions That Depend on Hypothesis Space Boundaries}
\label{sec-Tbeta}

Basing the ordering statistic on a score test can increase power over using simple Wald-type Z statistics  \citep[see][]{Chan:2003}.
Although this increased power has been shown in several simulation studies, it is not clear whether the increase is due to fewer ties for the score test, or from some other difference between the ordering statistics. A problem with the score statistic is that the induced ordering may change based on the $\beta_0$, since score statistics use $\beta_0$ in their calculation, whereas most other test statistics do not include $\beta_0$ in the calculation.
This can produce  non-coherence as was shown in Section~\ref{sec-IntuitionFailure} and Figure~\ref{figRohmelExample}.

Although the exact unconditional p-values and confidence intervals of this section can be
powerful, they are more difficult to calculate than the exact conditional ones described
in the next two sections: Section~\ref{sec-oneSidedCond} for p-values, and Section~\ref{sec-Melded} for compatible confidence intervals.

\section{One-Sided Conditional Exact Tests}
\label{sec-oneSidedCond}

\citet{Yate:1984}  argues that conditioning on the total number of failures is the proper strategy for this problem, and most of the discussants of the paper agreed with this (including Barnard, who first suggested the unconditional approach).  One of the main reasons that others had recommended the unconditional approach is an overemphasis on
the fixed significance level and the resulting power, which when used leads to more power for unconditional tests because the sample space has more values and hence is less discrete.
\citet{Yate:1984} argues (in his Section~9) that over reliance on the nominal significance level is not a good reason to prefer the unconditional test, and that p-values should be
reported instead of accept/reject decisions.  \citet{Yate:1984} also argues for conditioning on the total number of events ($X_1+X_2$), because that statistic is approximately ancillary to the effects of interest.
\citet{Cher:2004} quantifies the approximate ancillarity, showing that the absolute amount of information   ``is quite small unless [$\theta_1$ and $\theta_2$] are very far apart'', and  the proportion of information in the margins decreases with the sample size.
Recent reviews \citep[e.g.,][]{Lyde:2009}  have emphasized power arguments, and we review the choice of test from that perspective in Section~\ref{sec-power}. Historically, conditional tests have been important because of their much smaller computational burden compared to unconditional tests.
The computational burden for unconditional tests has become less important, although for some applications it may be a non-trivial concern  (e.g., big data applications
with small sample sizes but very many covariates being tested).

For the unconditional one-sided exact method, to calculate p-values we need to take the supremum of the probability that $T({\bf X})$ is at least as extreme than the observed
$T({\bf x})$ over   the parameter space $\Theta_0$ (see e.g., equation~\ref{eq:Tpvalues}).
This is a difficult calculation (see Section~\ref{sec-Computations}). An alternative method conditions on the sum $s=x_1+x_2$, and calculates the conditional probability.
The resulting conditional distribution is the extended hypergeometric distribution \citep{John:2005} also called Fisher's noncentral hypergeometric distribution \citep{Fog:2008}, which depends only on $\beta_{or}$. Additionally, because $s$ is fixed, we can write
the ordering function in terms of $X_2$ only. In fact, the only unique ordering function that makes sense and meets the BC conditions is $X_2$ itself (ordering on $n_1-X_1$ will be equivalent). So this simplifies the calculations if the effect measure is $\beta_{or}$.
For example, for testing $H_{0}: \beta_{or} \geq \beta_0$
use
\begin{eqnarray}
p_{Uc}({\bf x}, \beta_0) & = & \sup_{\theta \in \Theta_0} P_{\theta} \left[ T({\bf X}) \geq T({\bf x} ) | S \right]
=  \sup_{\beta_{or}:  \beta_{or} \geq \beta_0 } P_{\beta_{or}} \left[ X_2 \geq x_2  | S \right]  \nonumber \\
 & = &
 P_{\beta_{0}} \left[ X_2 \geq x_2 | S \right],  \label{eq:pUc}
\end{eqnarray}
where the last step follows because the conditional distribution is monotone in $\beta_{or}$ \citep{Meht:1985}.
The other conditional one-sided p-value, $p_{Lc}$ is calculated  similarly except by reversing the inequality.
These conditional p-values for testing $H_0: \beta_{or} =1$ (or equivalently $H_0: \theta_1=\theta_2$) are  Fisher's exact one-sided p-values.
We calculate the central confidence intervals on $\beta_{or}$ using equation~\ref{eq:onesidedCIs} except using the conditional exact
one-sided intervals instead of the unconditional ones.

\begin{figure} 
\vspace{6pc}
\includegraphics[width=5.5in]{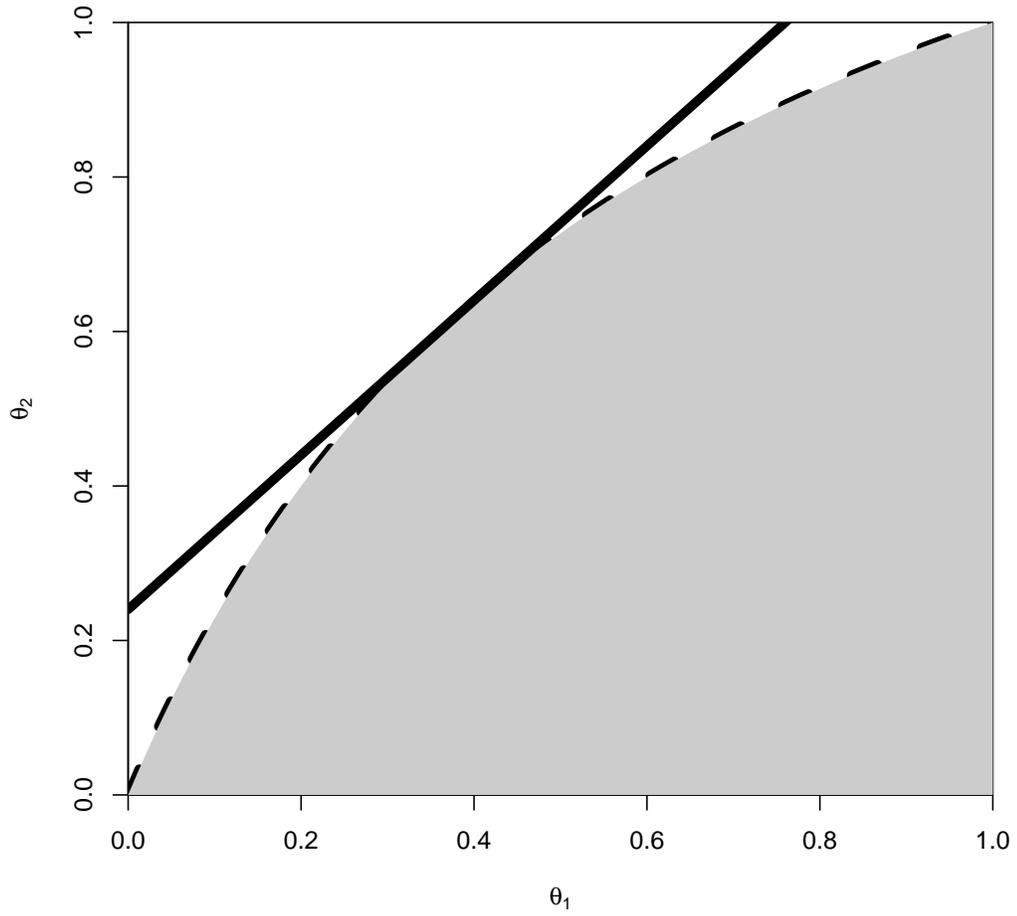}

\caption[]{97.5\% Confidence region based on one-sided conditional test of odds ratio (gray shaded area). Data is $x1/n_1=4/12$ and $x_2/n_2=8/15$. Upper 97.5\%
exact conditional limit on $\beta_{or}$ is $U=2.664$ (dotted line) and on $\beta_d$ is $U=0.240$  (solid line). The confidence region based on the upper limit for  $\beta_d$
is the gray region plus the white space between the dotted and solid line. We see that because the conditional probability depends on $\beta_{or}$ alone, that white space represents the lack of efficiency of basing the confidence region on $\beta_d$ instead of $\beta_{or}$.
}
\label{figcondCI}
\end{figure}

Now consider the other measures, $\beta_d$ and $\beta_r$.
At the boundary of equality, the one-sided hypotheses are equivalent. For example,
the following three null hypotheses give equivalent $\Theta_0$: (odds ratio) $H_{0U}: \beta_{or} \geq 1$, (ratio)  $H_{0U}: \beta_r \geq 1$,
and (difference)  $H_{0U}: \beta_d \geq 0$. Analogously for the other one-sided p-value.
But for boundaries not representing equality, $\Theta_0$ changes depending on the effect measure.
The simplification of the p-value calculation only works for the odds ratio.
For example, for the difference in proportions (i.e.,  $\beta=\beta_d$) there is no simplification analogous to  equation~\ref{eq:pUc}.
Figure~\ref{figcondCI} shows that the exact one-sided conditional confidence limit on $\beta_d$ is not efficient, because the  conditional distribution depends on $\beta_{or}$.
The upper $100(1-\alpha/2)\%$ limit for $\beta_d$, say $U_d$,   based on the upper limit for $\beta_{or}$, say $U_{or}$,
is \citep[see][Section 2]{Sant:1980}
\begin{eqnarray*}
U_d = \left\{
\begin{array}{cc}
0 &  \mbox{ if $U_{or} \leq 1$ } \\
\frac{ \sqrt{U_{or}} - 1}{\sqrt{U_{or}} + 1} & \mbox{ if $U_{or} > 1$ }
\end{array}  \right.
\end{eqnarray*}
There are better ways to get confidence intervals on $\beta_d$ and $\beta_r$ that provide
compatible inferences with the one-sided p-values with $\beta_0$ representing $\theta_1=\theta_2$.
We show these in the next section.

\section{Melded Confidence Intervals}
\label{sec-Melded}

\citet{FayP:2015} developed melded confidence intervals, a general method for creating confidence intervals for the two-sample case, that is closely related to the confidence distribution (CD) approach \citep{XieS:2013}.
Confidence distributions are a frequentist analog to the Bayesian posterior with a non-informative prior.
These melded confidence interval give compatible inferences with the central conditional tests.

Before discussing the binomial case, we consider the normal case because it is more straightforward.
Consider the difference in means between two normal samples with  different variances. Let $\mu_a$, $\bar{y}_a$, $n_a$,
and $s_a^2$ be, respectively, the mean, the sample mean, sample size,
and unbiased sample variance estimate for group $a$.
The two $100(1-\alpha/2)\%$ one-sided confidence intervals for the mean in group $a$, are
\begin{eqnarray*}
 \left( L_{\mu_a}(1-\alpha/2), \right. & \left. \infty \right)  &   \mbox{   and}  \\
   \left( -\infty, \right. & \left. U_{\mu_a}(1-\alpha/2) \right) &
\end{eqnarray*}
with
\begin{eqnarray}
L_{\mu_a}(1-\alpha/2) & = &  \bar{y}_a - F^{-1}_{n_a-1}(1- \alpha/2)  \frac{s_a}{\sqrt{n_a}}, \nonumber \\
&  \mbox{and} &  \label{normal.LU} \\
U_{\mu_a}(1-\alpha/2) & = &  \bar{y}_a + F^{-1}_{n_a-1}(1- \alpha/2)  \frac{s_a}{\sqrt{n_a}}, \nonumber
\end{eqnarray}
where $F^{-1}_{df}(q)$ is the $q$th quantile of the t-distribution with $df$ degrees of freedom.
The central $100(1-\alpha)\%$ confidence interval is the intersection of the two one-sided intervals,
\begin{eqnarray*}
\left(L_{\mu_a}(1-\alpha/2), U_{\mu_a}(1-\alpha/2) \right).
\end{eqnarray*}

The confidence distribution approach is a way to re-express the confidence interval.
Let $A$ and $B$ be two independent uniform $(0,1)$ random variables. Let
$M_{L_{\mu_a}} \sim L_{\mu_a}(A)$ and $M_{U_{\mu_a}} \sim U_{\mu_a}(B)$
be the lower and upper confidence distribution random variables for $\mu_a$,
where the randomness comes from $A$ and $B$, while $\bar{y}_a$ and $s_a/\sqrt{n_a}$ are treated as constants.
From (\ref{normal.LU}) and the probability integral transformation,  we re-express those random variables as
\begin{eqnarray*}
M_{L_{\mu_a}} & = &  \bar{y}_a - T_A  \frac{s_a}{\sqrt{n_a}}, \nonumber \\
&  \mbox{and} &  \label{normal.LU2} \\
M_{U_{\mu_a}} & = &  \bar{y}_a + T_B  \frac{s_a}{\sqrt{n_a}}, \nonumber
\end{eqnarray*}
where $T_A$ and $T_B$ are independent and distributed $t$ with $n_a-1$ degrees of freedom.
Because of the symmetry of the $t$ distribution about $0$, $M_{L_{\mu_a}}$ and $M_{U_{\mu_a}}$
have the same distribution, so in this case the lower and upper confidence distributions are equivalent, and we let  $M_{\mu_a} \equiv M_{L_{\mu_a}} \equiv M_{U_{\mu_a}}$
be the confidence distribution random variable associated with $\mu_a$.  In terms of the CD-RV,
the $100(1-\alpha)\%$ confidence interval for $\mu_a$ is
\begin{eqnarray*}
\left\{ q( \alpha/2, M_{\mu_a}),  q(1-\alpha/2, M_{\mu_a}) \right\}
\end{eqnarray*}
where $q(a,M)$ is the $a$th quantile of the random variable $M$.

The confidence distribution approach appears to be a confusing and roundabout way to express the confidence interval. The advantage comes when we want a confidence interval for $\mu_2- \mu_1$,
based on a two-sample problem with independent samples. Then we can write the $100(1-\alpha)\%$
confidence interval for $\mu_2-\mu_1$ as
\begin{eqnarray}
\left( q( \alpha/2, M_{\mu_2}-M_{\mu_1}),  q(1-\alpha/2, M_{\mu_2}-M_{\mu_1}) \right), \label{BFci}
\end{eqnarray}
which can be estimated with Monte Carlo simulation.
Expression~\ref{BFci} is equivalent to the Behrens-Fisher confidence interval,
and the confidence distribution approach gives a simple way to conceptualize it
\citep{FayP:2015}.  The traditional approach calculates the Behrens-Fisher statistic,
\begin{eqnarray*}
T_{BF} = \frac{\bar{y}_2 - \bar{y}_1}{ \sqrt{\frac{s_1^2}{n_1}+ \frac{s_2^2}{n_2}}}
\end{eqnarray*}
and calculates its
distribution, which depends on $n_1, n_2$ and $s_1/s_2$
\citep{KimC:1996}.

For the binomial problem the lower and upper confidence distributions are not equal.
Let $X_a \sim \mathrm{Binomial}(n_a, \theta_a)$, for $a=1,2$.
Let the $100(1-\alpha)\%$ exact central confidence interval  for $\theta_a$ (i.e., the Clopper-Pearson interval \citep{Clop:1934}) be
\begin{eqnarray*}
\left\{ L_{\theta_a}(1-\alpha/2), U_{\theta_a}(1-\alpha/2) \right\},
\end{eqnarray*}
where  $L_{\theta_a}(1-\alpha/2)$ and
$U_{\theta_a}(1-\alpha/2)$ are exact  $100(1-\alpha/2)\%$ one-sided confidence limits, for $\theta_a$ for $a=1,2$.
The lower and upper
CD random variables for group $a$ are $W_{La} = L_{\theta_a}( A_{a1})$
and $W_{Ua} = U_{\theta_a}( A_{a2})$, where $A_{ai}$ are independent uniform random variables.
This gives, $W_{La} \sim Beta( x_{a},n_a-x_a+1)$ with expectation $x_a/(n_a+1)$,
and $W_{Ua} \sim Beta( x_{a}+1,n_a-x_a)$ with expectation $(x_a+1)/(n_a+1)$, and using limits of parameters going to zero
we define $Beta(0,n+1)$ as a point mass at $0$ and $Beta(n+1,0)$ as a point mass at $1$.
The lower CD-RV is stochastically smaller than the upper CD-RV.
In CD form, the $100(1-\alpha)\%$ Clopper-Pearson interval is
\begin{eqnarray*}
\left\{ q( \alpha/2, W_{La}),  q(1-\alpha/2, W_{Ua}) \right\}.
\end{eqnarray*}
The $100(1-\alpha)\%$  melded confidence interval for $\theta_2-\theta_1$ is
\begin{eqnarray*}
\left\{ q( \alpha/2, W_{L2}-W_{U1}),  q(1-\alpha/2, W_{U2} - W_{L1}) \right\},
\end{eqnarray*}
where in order to be conservative for the lower limit,  we use the lower CD-RVs for $\theta_2$
but the upper CD-RV for $\theta_1$, and vice versa for the upper limit.
We can generalize this to other functions of $\theta = [\theta_1, \theta_2]$. Let $b(\theta)$
be a monotonic function of the parameters, such that
  $\beta = b(\theta)$ is increasing in $\theta_2$ and decreasing in $\theta_1$, within the allowable range of the parameters.
For the binomial problem all three parameters ($\beta_d$, $\beta_r$ and $\beta_{or}$) meet the monotonicity requirements, while for the normal two-sample problem the ratio of means (and odds ratios of means) does not meet those requirements. In general form, the $100(1-\alpha)\%$ (two-sided) melded confidence interval is given by
\begin{eqnarray*}
\left( q\left\{  \alpha/2, b([W_{U1},W_{L2}]) \right\}, q\left\{ 1-\alpha/2, b([W_{L1},W_{U2}]) \right\} \right).
\end{eqnarray*}
\citet{FayP:2015} conjectured that if the one-sample confidence interval procedures are valid, central, and nested, and $\beta=b(\theta)$
is increasing in $\theta_2$ for fixed $\theta_1$ and decreasing in $\theta_1$ for fixed $\theta_2$ (such as $\beta_d,$ $\beta_r$, and $\beta_{or}$), then the melded confidence interval is valid, nested and central. Some mathematical results,
simulations in several situations, and extensive numeric calculations in the binomial case supported this conjecture.
A rigorous proof of the conjecture is still needed.

Let $p_{Um}({\bf x}, \beta_0)$ and $p_{Lm}({\bf x}, \beta_0)$ be the one-sided melded p-values,
the p-values that match with the one-sided melded
confidence limits. Then for the binomial case, \citet{FayP:2015} showed that the one-sided  melded p-values equal the exact one-sided conditional p-values
when testing the null with margin $\beta_0$ which implies $\theta_1=\theta_2$.
For example, for testing $H_0: \beta_d \geq 0$, we have $p_{Um}({\bf x}, 0) = p_{Uc}({\bf x}, 0)$,
and for testing $H_0: \beta_r \geq 1$, we have $p_{Um}({\bf x}, 1) = p_{Uc}({\bf x}, 1)$.
Because the melded confidence intervals are nested, by Theorem~\ref{THM} the melded confidence intervals are compatible with the p-values from the one-sided Fisher's exact test.

\begin{figure} 
\vspace{6pc}
\includegraphics[width=5.5in]{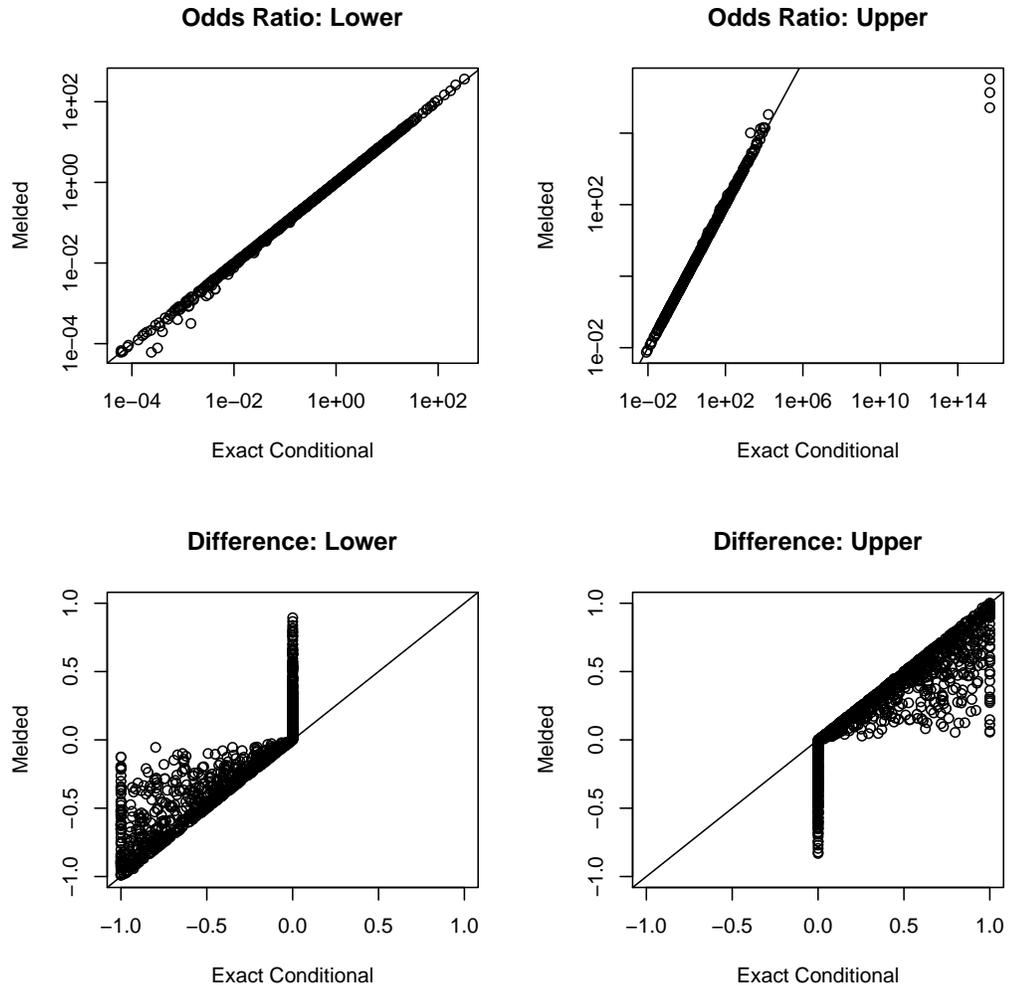}

\caption[]{Lower and Upper limits associated with 95\% central confidence intervals by exact conditional method and melding method.
Simulated data where $n_a$ is simulated from uniform on $1$ to $100$, and $x_a$ is uniform on $0$ to $n_a$, 1000 replications.
Calculation used the {\sf exact2x2} R package for melded confidence limits and {\sf fisher.test} from the {\sf stats} package for the exact conditional limits.
The limits for $\beta_{or}$ agree well; the identity line for the upper limits does not reach the top right corner of the graph because of some extreme data (e.g., $x_1/n_1=1/68$ and $x_2/n_2=57/61$) perhaps caused by numeric issues in the computation. In contrast, the limits for $\beta_d$ show that the
melded are shorter intervals (lower is larger, upper is smaller).
}
\label{figCompareCondMeld}
\end{figure}

The melded CIs for $\beta_{or}$ are very close  to the exact conditional ones, but the melded CIs for $\beta_d$ are more efficient (lower are larger, and upper are smaller)
than the exact conditional ones (see Figure~\ref{figCompareCondMeld}).

\section{Non-central Confidence Intervals and Associated Tests}
\label{sec-twoSidedTests}

Let $T_{ts}({\bf x}) \equiv T_{ts}({\bf x}, \alpha, \beta_0)$ be an ordering function
for testing the two-sided null $H_{0}: \beta=\beta_0$, with smaller values suggesting $\beta$ further away from the null.
We can create exact unconditional two-sided p-values using
\begin{eqnarray*}
p_{ts}({\bf x}, \beta_0) & = & \sup_{\theta \in \Theta(\beta_0)} P_{\theta} \left[ T_{ts}({\bf X}) \leq T_{ts}({\bf x}) \right]
\end{eqnarray*}
and exact conditional two-sided p-values using
\begin{eqnarray*}
p_{ts}({\bf x}, \beta_0) & = & \sup_{\theta \in \Theta(\beta_0)} P_{\theta} \left[ T_{ts}({\bf X}) \leq T_{ts}({\bf x}) | S=s \right].
\end{eqnarray*}
which simplifies to
\begin{eqnarray}
& & p_{ts}({\bf x}, \beta_0)  =  P_{\beta_0} \left[ T_{ts}({\bf X}) \leq T_{ts}({\bf x}) | S=s \right], \label{eq:pvalTSCond}
\end{eqnarray}
if $\beta=\beta_{or}$.

For example, consider $T_{ts}({\bf x}, \beta_0) = f({\bf x}, \beta_0)$, where $f$ is the probability mass function for the extended hypergeometric distribution with parameter $\beta_{or}=\beta_0$. The associated exact conditional p-value when $\beta_0=1$ is the usual Fisher's exact test, called the {\it Fisher-Irwin test} since it was proposed by \citet{Irwi:1935}
  and to distinguish it from the central Fisher's exact test created by doubling the minimum of the one-sided Fisher's exact p-values.
Using Fisher's exact p-values (either Fisher-Irwin or central version) as an ordering function in an unconditional exact test gives a version of Boschloo's  test.
\citet{Bosc:1970} showed that using the Fisher-Irwin p-values in this way is uniformly more powerful than  the Fisher-Irwin test.
In an analogous way,
Using either one-sided or central Fisher's exact p-values as ordering functions in an unconditional test,
is also uniformly more powerful than the original Fisher's exact versions of those tests \citep{Lyde:2009}.

\citet{Blak:2000} studied non-central confidence sets that always are subsets of the central confidence sets in one parameter distributions.
To translate into this problem, we consider only the conditional distribution based on $S=s$ and $\beta=\beta_{or}$.
Start with $T({\bf x})=x_2$, a one-sided ordering function for the conditional problem (see Section~\ref{sec-oneSidedCond}).
Define
\begin{eqnarray*}
 \gamma({\bf x},\beta) =  \min  \left\{ P_{\beta}[X_2 \leq x_2 | S=s ],
  P_{\beta}[X_2 \geq x_2 | S=s ]  \right\}.
\end{eqnarray*}
Let the two-sided ordering function for Blaker's test be
\begin{eqnarray*}
T_{B}({\bf x},\beta) = P_{\beta} \left[ \gamma({\bf X}, \beta) \leq  \gamma({\bf x}, \beta) | S=s \right].
\end{eqnarray*}
{\it Blaker's test} two-sided p-value is $p_{B}({\bf x},\beta_0)$ from equation~\ref{eq:pvalTSCond} using $T_{ts}=T_{B}$,
and the associated $100(1-\alpha)\%$ confidence region is
\[
C_{B}({\bf x},1-\alpha) = \left\{ \beta: p_{B}({\bf x}, \beta) > \alpha \right\}.
\]
\citet{Blak:2000} showed that this gives smaller confidence sets than the central CIs.
Specifically, $C_{B}({\bf x},1-\alpha) \subset C_c({\bf x},1-\alpha)$, where $C_c$ is the exact conditional central CI using
the one-sided ordering function $T({\bf x})=x_2$.
Let the $100(1-\alpha)\%$ matching confidence interval to $p_{B}$ be the smallest interval that
contains $C_{B}$.

Consider the conditional two-sided tests for $\beta_0=1$ when $x_1/n_1=8/14$ and $x_2/n_2=1/7$.
Conditionally on $X_1+X_2=9$, the support of $X_2$  is $\left\{ 0, 1, \ldots, 7 \right\}$.
In Table~\ref{tabBFI} we give the values of $f({\bf x},1)$, $\gamma({\bf x}, 1)$,
and $T_B({\bf x},1)$. Note, $f([8,1],1)=f([4,5],1)$.
Suppressing the $\beta_0=1$ term in the functions, the p-value for the Fisher-Irwin test is,
\begin{eqnarray*}
p_{FI}([8,1]) & = & f([9,0]) + f([8,1]) + f([4,5])+f([3,6]) + f([2,7]) = 0.159,
\end{eqnarray*}
for the Blaker test is
\begin{eqnarray*}
p_{B}([8,1]) & = & f([9,0]) + f([8,1]) +f([3,6]) + f([2,7]) = 0.087,
\end{eqnarray*}
and for the central Fisher's exact test is
\begin{eqnarray*}
p_{c}([8,1]) & = & 2* \left\{ f([9,0]) + f([8,1]) \right\} = 0.157.
\end{eqnarray*}
This example was chosen to clarify the differences between the tests, but often the Fisher-Irwin and Blaker tests give the same p-values.
The calculation of the matching 95\% confidence intervals involves calculating a series of p-value functions for changing $\beta_0$,
which may not be unimodal for Blaker's test or the Fisher-Irwin test (see e.g., Figure~\ref{figFisherIrwin}),
so the algorithm is not simple \citep{Fay:2010}. The 95\% confidence intervals are: $C_{FI}({\bf x}, 0.95) =(0.005, 1.53)$,
$C_B({\bf x}, 0.95) =(0.005, 1.53)$,  and $C_c({\bf x}, 0.95) =(0.002, 1.62)$.
The original (i.e., two-sided) CSM test of \citet{Barn:1947} is more difficult to calculate (see Appendix~\ref{sec-BarnardCSM}), it gives a
two-sided p-value of $0.089$, and we know of no software to calculate the matching confidence interval.

\begin{table}[ht]
\caption{The hypergeometric probability mass function ($f$), $ \gamma({\bf x},1)$ function,
and the two-sided ordering function for Blaker's exact test ($T_{B}({\bf x},1)$), for $x_1/n_1=8/14$ and $x_2/n_2=1/7$.
\label{tabBFI}
 }
\centering
\begin{tabular}{rrrrrrrrr}
  \hline
 & $x_2=0$ & $x_2=1$ & $x_2=2$ & $x_2=3$ & $x_2=4$ & $x_2=5$ & $x_2=6$ & $x_2=7$ \\
  \hline
$f({\bf x},1)$ & 0.007 & 0.072 & 0.245 & 0.358 & 0.238 & 0.072 & 0.009 & 0.000 \\
$\gamma({\bf x},1)$ & 0.007 & 0.078 & 0.324 & 0.676 & 0.319 & 0.080 & 0.009 & 0.000 \\
$T_B({\bf x},1)$ & 0.007 & 0.087 & 0.642 & 1.000 & 0.397 & 0.159 & 0.016 & 0.000 \\
   \hline
\end{tabular}
\end{table}


\citet{Agre:2001} showed that to create two-sided CIs with shorter expected length,
it is generally better to invert p-values from two-sided hypothesis tests
that are not central. This makes sense because centrality is a restriction, and two-sided tests without that restriction will leave room
for improving expected CI length. For the two-sample binomial problem, basing $T_{ts}({\bf x}, \beta_0)$ on score tests gives good expected CI length;
see \citet{Chan:1999} for $\beta_d$ and \citet{Agre:2002} for $\beta_{or}$.
Despite this apparent improvement, if directional inferences are needed, then central confidence intervals
are recommended (see Section~\ref{sec-3sided}).

\section{Mid-p Methods: Improving Accuracy by Sacrificing Validity}
\label{sec-midp}

The mid-p value is a modification of a p-value for discrete data.
Instead of calculating the probability of
observing equal or more extreme responses, the mid-p value is $0.5$ times the probability of equality plus the probability of more extreme. For example,
the conditional exact p-value of equation~\ref{eq:pUc} becomes
\begin{eqnarray}
p_{Uc-mid}({\bf x}, \beta_0) & = & P_{\beta_0} \left[ X_2 > x_2 | S \right] + \frac{1}{2}  P_{\beta_0} \left[ X_2 = x_2 | S \right]. \label{eq:pcumid}
\end{eqnarray}
\citet{Hwan:2001} gave some optimality criteria for the mid-p approach applied to one parameter situations, which applies to the conditional
test using  $\beta_{or}$ since the conditional probability is completely described by only the $\beta_{or}$ parameter.
They show that for one-sided or two-sided hypothesis tests, the loss based on squared error between an indicator that $\beta \in \left\{ b(\theta): \theta \in \Theta_0 \right\}$
and the p-value function, and shows that for all $\beta \in \left\{ b(\theta): \theta \in \Theta_1 \right\}$ (and $\beta=\beta_0$) the expected loss is less than or equal to (strictly less than) the expected loss
from any randomized exact p-value function (Theorem 3.3 and 4.3 with \citet{Yang:2004}).
\citet{Fell:2010} showed minimaxity under squared error and linear loss, and also showed that of all non-randomized ordered decision rules, the mid-p version
is the only one that has expectation $1/2$ under a point null hypothesis.

\section{Computational Issues}
\label{sec-Computations}

Overall, conditional p-values are much easier to calculate than unconditional ones, since they do not require taking the supremum over the null space.
The melded confidence intervals allow matching CIs to conditional tests of $\theta_1=\theta_2$, and are very quick to calculate, since they use numeric integration.
There may be some precision issues in the numeric integration for extreme data sets.

The main computational speed issues apply to unconditional tests, since they require computing the supremum.
\citet[][p. 161]{Rohm:1999}  showed that for ordering statistics, $T$, that meet the BC conditions,
the supremum in the p-value calculation is on the boundary between hypotheses. For example,
\begin{eqnarray*}
& & \sup_{\theta \in \Theta_0}
 P_{\theta} \left[ T({\bf X}) \geq T({\bf x}) \right] =
\sup_{\theta: b(\theta)=\beta_0} P_{\theta} \left[ T({\bf X}) \geq T({\bf x}) \right].
\end{eqnarray*}
For example, the score statistic on $\beta_d$ \citep{Farr:1990} has been shown to follow the BC conditions for fixed $\beta_0$ \citep{Rohm:2005}.
Further, if  $T$ meets the BC conditions and does not depend on $\beta_0$, then Theorem 3.1 of \citet{Kaba:2005} shows that the exact unconditional one-sided p-values based on $T$
are either nonincreasing (for $p_U({\bf x}, \beta_0)$) or nondecreasing (for $p_L({\bf x},\beta_0$)) in $\beta_0$ for fixed ${\bf x}$.
This property means that for these p-values, the associated $100(1-\alpha/2)$ one-sided confidence intervals can be easily calculated by finding the
value $\beta_0$ where the p-value equals $\alpha/2$.

Calculation using Barnard's CSM p-value ordering can be very slow, because determining the ordering itself requires p-value calculation.
\citet{Rohm:2013} discussed one-sided exact unconditional tests using Barnard's CSM p-value ordering, except with  breaking ties in a manner that does not worry about symmetry equivariance.
They also do not  worry about the exact ordering for very small p-values. This can speed up the calculations substantially.

Table~\ref{tab-methods}  reviews different methods, their properties of centrality and compatible inferences,
 and approximate ranking of computational speed and power. The last column gives some software availability for the methods;
 it is not a comprehensive list, and only considers SAS 9.4, R (with packages), and StatXact 11.

\clearpage

\begin{landscape}

\begin{table}
\caption{ Valid (and Mid-p adjusted) Methods for Two-Sample Binomial Problem, and
some Properties, References, and Software \label{tab-methods} }

{\tiny
\begin{tabular}{cccccccc}

Method & Central & Compat.    & Comput. & Power/  &  References & Sect. & Software$^{***}$ \\
       &         & Infer.  & Speed$^{*}$  &      Efficiency$^{**}$   &        &          &  \\ \hline \hline
Smallest CI & yes & no    & 3  &  1   & \citet{Wang:2010} (for $\beta_d$)  & \ref{sec-Talpha} & Rpkg:ExactCIdiff (for $\beta_d$ CI only) \\
            &     &       &    &       & \citet{Wang:2015} (for $\beta_r, \beta_{or}$) &  &
\\ \hline
Barnard's CSM &   both & ? & 3 & 1 & \citet{Barn:1947} & \ref{sec-UncondBasic} & Rpkg: Exact(p-value only) \\ \hline
Boschloo Test &  both &  both  & 2 & 2 & \citet{Bosc:1970} & \ref{sec-twoSidedTests} & Rpkg: Exact (p-value only), exact2x2 \\ \hline
Uncond Exact   &  no & no & 2  &  2   &  \citet{Chan:1999} (for $\beta_d$) & \ref{sec-twoSidedTests} & StatXact-11 (only $\beta_d, \beta_r$),  \\
Score Stat     &     &    &    &      &   \citet{Agre:2001} (for $\beta_d, \beta_r$)                & & SAS 9.4 (only $\beta_d, \beta_r$),         \\
(square $T$)   &     &    &     &     &   \citet{Agre:2002} (for $\beta_{or}$)                      & &  Rpkg: exact2x2 (tsmethod=``square'')    \\ \hline
Uncond Exact   &  yes   & yes &  2 & 2($\beta_d$) &                                                          & \ref{sec-refinement} & Rpkg: exact2x2 (tsmethod=``central'')    \\
$\beta$ Estimates &     &      &   & 5($\beta_r$)  &                                                   &    \ref{sec-orderingRatio}                 &                     \\
with tie break    &     &      &    & 5($\beta_{or}$)  &                                                        &     &                     \\ \hline
Uncond Exact   &  no  &  no   &    2  &  2  & \citet{Mehr:2003}                                      & \ref{sec-twoSidedTests}  & StatXact-11  (only $\beta_d$) \\
Wald Stat ($T^2$)     &     &      &       &     &                                                        &   &  Rpkg: exact2x2 (tsmethod=``square'')         \\ \hline
Uncond Exact   &  yes   & yes &  2 & 3($\beta_d$) &        \citet{Barn:1945}                                & \ref{sec-UncondBasic} & Rpkg: exact2x2 (tsmethod=``central'')    \\
$\beta$ Estimates &     &      &   & 5($\beta_r$, $\beta_{or}$)   &        \citet{Mehr:2003}                                                & \ref{sec-orderingRatio}     &                     \\ \hline
Cond Exact with & no  &  no     &  1     &   3   & \citet{Fish:1934} (for p-value)                          &  \ref{sec-oneSidedCond}     & Rpkg:exact2x2    \\
Fisher-Irwin    &     &         &        &        &    \citet{Fay:2010} (for CI)                              &             &                  \\
Exact Test      &     &         &        &        &                                                  & &                  \\ \hline
Cond Exact with &  no  & no  & 1  & 3  &  \citet{Blak:2000}                                        & \ref{sec-twoSidedTests} & Rpkg: exact2x2  \\
Blaker Method   &        &      &    &   &  \citet{Fay:2010}                                       &  &                   \\ \hline
Cond Exact with & yes    & yes   &  1   &  4    & \citet{Fish:1934} (for p-value)                  & \ref{sec-Melded} & Rpkg: exact2x2      \\
Melded CIs      &        &       &      &        &  \citet{FayP:2015}   (for CI)                            &  &                     \\  \hline
Cond exact with & yes    & yes   & 1    &  4     & \citet{Agre:2001}                               & \ref{sec-oneSidedCond} & SAS 9.4 (use double one-sided  \\
tail approach CI &       &      &       &        &  \citet{Fay:2010}                               &  & Fisher's exact p-values)       \\
(only for $\beta_{or}$) &  &     &      &        &                                                 &   & StatXact-11, Rpkg: exact2x2 \\ \hline \hline
 & \mbox{} \\
Adjustment & \multicolumn{5}{c}{Notes} & Sect. & Software  \\  \hline
Berger-Boos & \multicolumn{5}{l}{Adjustment by \citet{Berg:1994} applies to unconditional exact tests} & \ref{sec-otherImprovements} & StatXact-11, Rpkg: exact2x2 \\
          & \multicolumn{5}{l}{and generally increases power }  &   & Rpkg: Exact(p-values only) \\ \hline
E+M  & \multicolumn{5}{l}{Adjustment by \citet{Lloy:2008} applies to unconditional exact tests} & \ref{sec-otherImprovements} &  Rpkg: exact2x2            \\
 &  \multicolumn{5}{l}{ and generally increases power} \\ \hline
Mid-p & \multicolumn{5}{l}{Applies to any method, increases power at the cost of validity}                                         & \ref{sec-midp} &  Rpkg: exact2x2   \\
      & \multicolumn{5}{l}{ }                                                                                               &  &  SAS 9.4 (not all tests)       \\ \hline \hline
\end{tabular} \\
$^*$ Approximate computation speed:  1=fast, 2=moderate, 3=slow.
$^{**}$ Approximate power/efficiency: 1=higher power/shorter CI, $\ldots$, 5=lower power/larger CI. \\
$^{***}$ Software (not comprehensive, only considered R, SAS and StatXact): R packages available at \url{https://cran.r-project.org/}. \\
 For SAS the methods are available in PROC FREQ using exact option. The value ``both''  denotes there could be versions with and without the property,
 and ``?'' denotes that it is not clear if the matching confidence intervals are compatible with the p-values because confidence intervals have not been studied with that test (although it
 is likely the method will not be compatible because it is similar to the smallest CI method).

}
\end{table}

\end{landscape}

\section{Power and Efficiency Comparisons}
\label{sec-power}

A comprehensive simulation or calculation comparing different methods with respect to power or efficiency  is beyond the scope of this review.
Here we review a few of the best of those types of papers and add an example and some graphical calculation results to supplement
the previous literature on the topic. In essence this section gives some detailed justification for the rough power/efficiency classifications
listed in Table~\ref{tab-methods}.

In general conditional tests (e.g., Fisher's exact tests) are less powerful than the best of the unconditional tests,
 because the latter tests are less discrete
\citep{Lyde:2009}.
\citet{Mart:1994} provide a very comprehensive power comparison of several valid unconditional tests
(including tests based on either an  ordering function of the difference in sample proportions, or on some test-based ordering functions related to Fisher's exact p-value, the unpooled Z test, or Barnard's CSM test).
They only considered ordering functions that do not depend on $\alpha$ or $\beta_0$ (since they only consider power
to show $\theta_2 > \theta_1$ [i.e., with $\beta_0 = 0$ for the difference  or $\beta_0=1$ for the ratio or odds
ratio] the ordering functions automatically do not depend on $\beta_0$).
\citet{Mart:1994} based power comparisons on expected power assuming bivariate uniformly distributed $(\theta_1,\theta_2)$.
They found that Barnard's CSM test was the most powerful on average, and that ordering by
either the unpooled $Z$ statistics for the difference in means or Fisher's exact p-values (i.e., a Boschloo-type test) gave
the next best power. \citet{Mart:1994} did not include a pooled Z test, but \citet{Mehr:2003} did, and they showed that
the pooled Z test can have much better power with unequal sample sizes.  So in general we can recommend ordering by the pooled Z instead of the unpooled Z.
Since Barndard's CSM test is difficult to calculate, \citet{Mart:2002} compared many approximations to that value.
They concluded that the mid-p Fisher's p-value was the best approximation to the CSM test, although it could be conservative for very small samples.
\citet{Hirj:1991} did extensive calculations finding the type I error rate for the exact conditional mid-p one-sided and two-sided (Fisher-Irwin-type) tests.
They found that out of 3125 sample size and parameter situations (all with $\theta_1=\theta_2$), typically 90-95\% of both types of the mid-p p-value
when used to test at a 5\% significance level, had type I error rates less than or equal to 5\%.  Further, \citet{Lyde:2009} stated that the mid-p version of
the Fisher-Irwin test approximates the Fisher-Boschloo test well, and the latter test (or the exact unconditional test on Pearson's chi-squared test) was their recommendation.


For confidence intervals, we focus on two papers.
\citet{Chan:1999} compared unconditional confidence intervals based on
estimates or tests on the difference: the difference in proportions, the unpooled Z statistic, the score statistic
(which they called the $\delta$-Projected Z statistic), and the likelihood ratio statistic. They tried all with and without the \citet{Berg:1994} adjustment. They showed
the score statistic with no adjustment generally gave shorter expected confidence interval length.
\citet{Sant:2007} did a very comprehensive set of calculations for $\beta_d$ confidence intervals,
calculating expected coverage and confidence interval length
for a $100 \times 100$ grid of values of $(\theta_1, \theta_2)$.
They compared three valid methods and two approximate methods, including the unconditional method based on a two-sided score test,
the unconditional method based on two one-sided score tests, and an approximate method of \citet{CoeT:1993}.
The results show that of the valid methods, the  unconditional method based on the two-sided score
test statistic had the lowest expected length, while the central unconditional method based on two one-sided score tests
had larger expected length. However, if directional inferences are important, then the proper comparison should be
the former method using $100(1-2\alpha)\%$ intervals compared to the latter method using $100(1-\alpha)\%$ intervals (see Section~\ref{sec-3sided}).
Further, score tests may lack coherence (see Figure~\ref{figRohmelExample}).
\citet{Sant:2007}  recommended the approximate method of \citet{CoeT:1993}, which had shorter expected length confidence intervals and
gave coverage above the nominal except in less than $0.6\%$ of the cases.
\citet{Fage:2015} also recommends for small samples the exact unconditional confidence intervals with the ordering function the
two-sided score test statistic. \citet{Fage:2015} mentions using one-sided tests if direction is important.

We now compare  score tests to other tests not included in the previous simulations.
Between unconditional tests applied to $\beta_r$ and $\beta_{or}$, the ordering based on score tests or the ordering based on one-sided mid-p Fisher's exact p-values \cite{Mart:1998}  perform much better than ordering by estimates with tie breaks as in Section~\ref{sec-orderingRatio}.
For example, with $n_1=n_2=20$, $\theta_1=0.4$, $\theta_2=0.8$, and a one-sided $0.025$ significance level,
power is  73\% for score-based or mid-p Fisher-based tests of both $\beta_r$ and $\beta_{or}$ but is very small for the test that orders by estimates with tie breaks (power $\approx 0$ for $\beta_r$ and power $ \approx 1\%$ for $\beta_{or}$). Power increases slightly for the latter tests with a Berger and Boos adjustment and $\gamma=10^{-6}$
(power is 11\% for $\beta_r$ and 16\% for $\beta_{or}$).
In contrast, for $\beta_d$ in that example all three methods of ordering with or without the Berger-Boos adjustment give 73\% power.

\begin{figure} 
\vspace{6pc}
\includegraphics[width=4.0in]{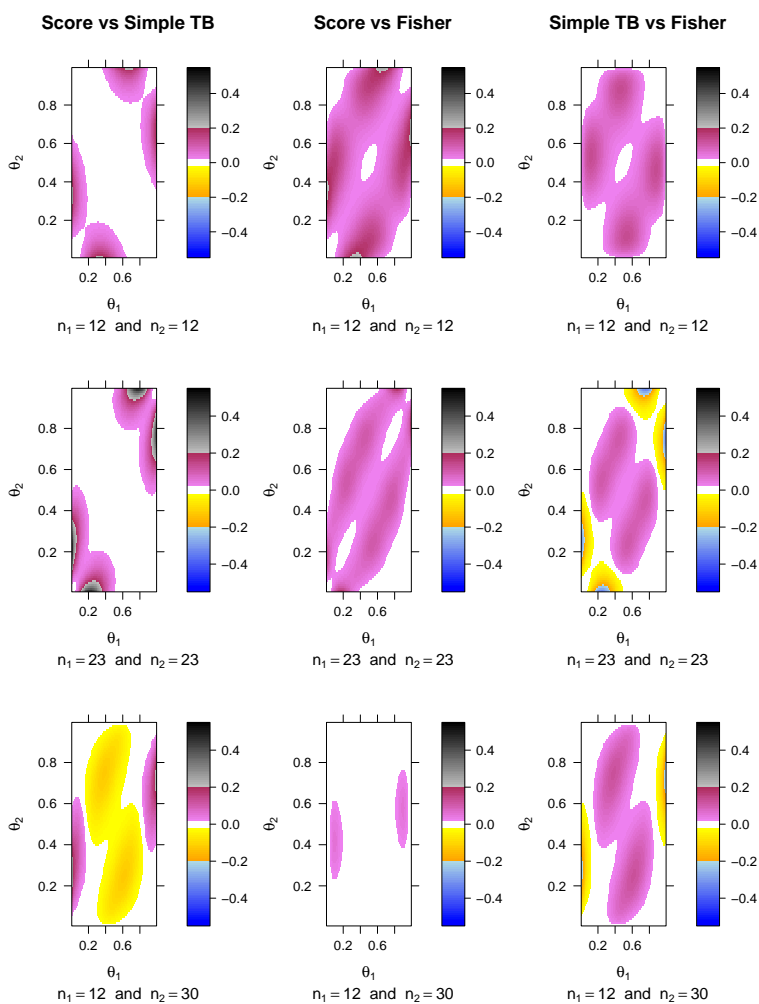}

\caption[]{
Comparison of powers  for testing $\theta_1=\theta_2$ using central tests at the two-sided $0.05$ level.
The three tests compared are ``score''= unconditional exact test based on
 the score test of the difference in proportions, ``simple TB''= unconditional exact test based on the difference in proportions using a simple tie-break (see Section~\ref{sec-refinement}), and ``Fisher''= tests based on central Fisher's exact test.
For columns labeled Test 1 vs Test 2, the result is power of Test 1 minus Power of Test 2, so that positive values (pink and gray) indicate that
Test 1 is more powerful.   White indicates that powers are within $0.025$ of each other.
Colors are smoothly changing so that light blue (-0.2 to -0.3) are next to yellow and gray (0.2 to 0.30) is next to pink.
}
\label{fig:power}
\end{figure}

Figure~\ref{fig:power} compares powers on the two-sided $0.05$ level central tests that $\beta_d=0$.
Powers are calculated on a  $99 \times 99$ grid of values of $(\theta_1,\theta_2)$.
We plot the difference in powers between all pairs of three tests: two unconditional exact tests (one based on the score test for the difference in proportions,
 and one based on the difference in proportions with a tie break) and the conditional test (the  central Fisher's exact test).
 We find, as expected, that unconditional tests do better, and that the simple method with a tie break does well when the sample sizes are not equal
   \citep[see e.g.,][for  a different set of simulations showing a similar result for the two-sided test]{Mehr:2003}.

\begin{figure} 
\vspace{6pc}
\includegraphics[width=4.0in]{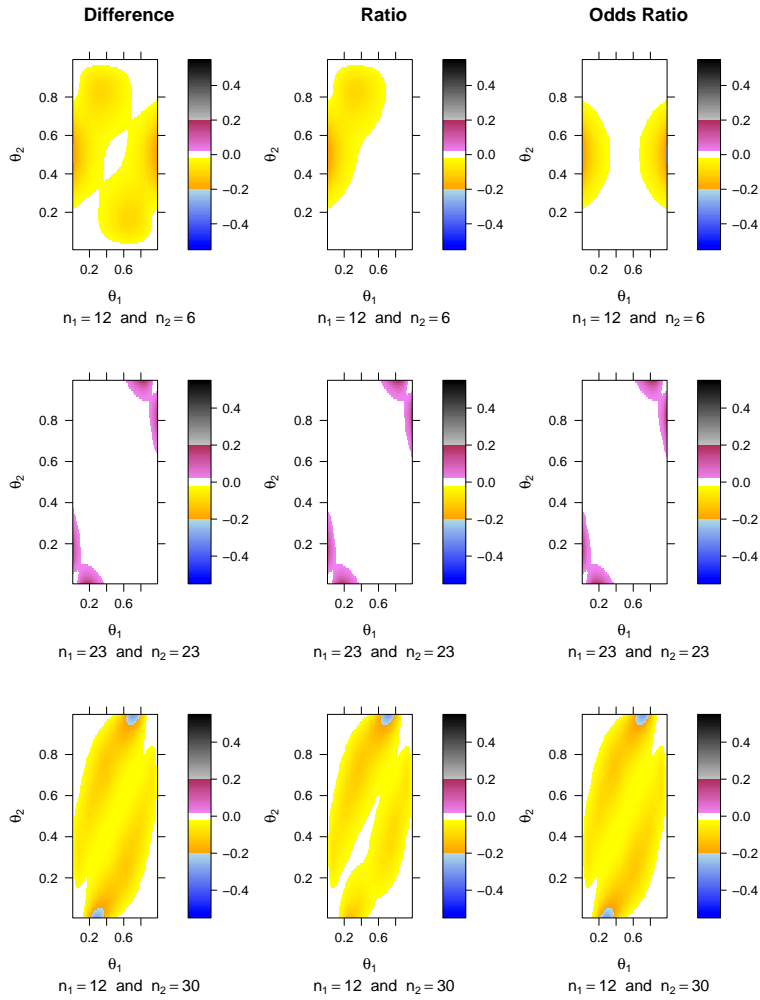}

\caption[]{
Power of unconditional exact score test  minus power of unconditional exact test based on ordering by Fisher's exact test one-sided mid p-value.
Negative values (yellow and blue) denote parameter values in which the latter test is more powerful.
The unconditional exact score tests are defined based on testing either $H_0: \beta_d=0$ (first column), $H_0: \beta_{or}=1$ (second column), or
$H_0: \beta_r=1$ (third column).
White indicates that the two powers are within $0.025$ of each other.
Colors are smoothly changing so that light blue (-0.2 to -0.3) are next to yellow and gray (0.2 to 0.30) is next to pink.
Additional calculations with $n_1=12, n_2=12$ showed nearly equal powers  (all white) for all three columns and are not plotted.
}
\label{fig:power2}
\end{figure}

Figure~\ref{fig:power2} compares unconditional exact tests ordered by score statistics (on either $\beta_d=0, \beta_{or}=1,$ or $\beta_r=1$) compared to
 unconditional exact tests based on the mid p-values from the one-sided Fisher's exact test. We find that the latter tests are generally more powerful.

\section{Recommendations}
\label{sec-Recommendations}

There are many ways to perform frequentist inferences on the two-sample binomial
problem. Our extensive review focused on valid inferences and highlighted
practical properties of tests.
We give a few recommendations.
\begin{enumerate}
\item
Use central confidence intervals with either a central p-value, or the minimum of the one-sided p-values.
Using non-central two-sided CIs can slightly decrease expected CI length, but at a cost in terms of allowable one-sided inferences.
Since we  usually care about the direction of effect, non-central CIs are not routinely recommended.
\item Avoid maximizing power or minimizing the expected length of the confidence interval, because it increases computational burden and can lead to incoherent p-values and non-nested CIs.
\item For fast calculations use one-sided conditional exact tests and melded confidence intervals.
\item For more power use unconditional one-sided valid p-values and associated central CIs. For inferences on $\beta_d$, order based on the
difference in sample proportions, except break ties while maintaining the
BC conditions, and do not let the ordering function depend on $\beta_0$ or $\alpha$. This will ensure monotonicity of p-values as a function of $\beta_0$, allowing for relatively fast calculations, while preserving coherence and nestedness.
For inferences on $\beta_r$ and $\beta_{or}$, using the simple function with a tie breaking ordering has much
smaller power than the score method or ordering based on one-sided mid-p Fisher's exact p-values. The score method causes incoherence or non-nestedness,
while the mid-p Fisher p-value ordering does not. Because the latter method only uses the mid p-values for ordering within the exact unconditional test framework,
the resulting p-values are valid. Further, for inferences on $\beta_d$, the mid-p ordering meets the BC conditions and is relatively fast to calculate.
\item If validity is not vital, then the mid-p conditional tests are a good approximation to the more powerful of the unconditional exact ones.
Additionally, with a large proportion of situations with $\theta_1=\theta_2$, the mid-p conditional tests still have type I error rates less than the nominal value.
\end{enumerate}

\appendix

\section{Proof of Theorem~\ref{THM}}
\label{sec-proof}

\begin{description}
\item[Proof of statement~\ref{thmUCR}]:
\begin{description}
\item[(Compatible Inferences) $\Rightarrow$ ($C_I=C$):]  If the confidence region associated with a p-value is not an interval, then there must be an $\alpha$ and $\beta_0$ such that
$p({\bf x}, \beta_0) \leq \alpha$ and $\beta_0 \in C_I({\bf x}, 1- \alpha)$, which contradicts the compatible inferences, therefore $C_I({\bf x},1-\alpha) = C({\bf x}, 1-\alpha)$.
\item[($C_I=C$)  $\Rightarrow$ (Compatible Inferences):] If the confidence region associated with the p-value is the matching confidence interval, then the inferences
are compatible by definition (equation~\ref{eq:CIbyp}).
\end{description}
\item[Proof of statement~\ref{thmUN}, (Compatible Inferences) $\Rightarrow$ (Nested CI):]
We show the contrapostive.
If a method has non-nested CIs, then there exists some $\alpha_1 < \alpha_2$ and some $\beta_0$ such that
$\beta_0 \notin C_I({\bf x},1-\alpha_1)$ and $\beta_0 \in C_I({\bf x},1-\alpha_2)$.  If the method had compatible inferences, then $p({\bf x},\beta_0) \equiv p  \leq \alpha_1$
and $p > \alpha_2$. This leads to the contradiction, $p \leq \alpha_1 < \alpha_2 < p$, so the method must not have compatible inferences, and
we have proven the result.
\item [Proof of statement~\ref{thmUOS},
(Compatible Inferences) $\Rightarrow$ (Coherence):]
 From statement~\ref{thmUN},  compatible inferences imply nested CIs. For one-sided p-values,  compatible inferences with
 nested CIs imply that the p-values are non-decreasing as the null space expands (e.g., $\beta_0$ gets larger when $H_0: \beta \leq \beta_0$), and hence are coherent by definition.
 For two-sided p-values, because of compatible inferences and nested CIs, the p-values are increasing (i.e., non-decreasing) as $1-\alpha$ decreases. This is directional coherence by definition.
\end{description}

\section{Barnard's CSM Ordering}
\label{sec-BarnardCSM}

Because \citet{Barn:1947} defined his CSM ordering as a two-sided ordering, there may be more than one way to generalize the idea to a one-sided ordering.
We present two one-sided orderings here \citep[see][for alternative algorithmic details]{Silv:1997}.
Consider first the bottom-up CSM ordering, where we start with the point with the lowest value of
$T({\bf x})$, which is $[x_1,x_2] = [n_1, 0]$, and make that the first rejection region, say $R_1$, and let $T([n_1,0])=1$.
Then repeat the following algorithm to create the $j$th rejection region (for $j=2,\ldots$) until all points have been ordered:
\begin{enumerate}
\item Let $Q_{j}$ be the set of points such that when each individual point is added to $R_{j-1}$, the resulting set meets the BC condition.
  Let the elements of $Q_{j}$ be $\left\{ {\bf x}_{j1}, \ldots, {\bf x}_{jm_j} \right\}$.
  For example, when $j=1$, then $Q_{2} = \left\{ [n_1,1], [n_1-1,0] \right\}$, with ${\bf x}_{21}=[n_1,1]$ and ${\bf x}_{22}=[n_1-1,0]$.
\item Calculate $p_U({\bf x}_{jk})$ for each member of $Q_{j}$, where
\begin{eqnarray*}
p_U({\bf x}_{jk}) & = &  \sup_{\theta: \theta_2 = \theta_1} P_{\theta} \left[ T({\bf X}) \leq T({\bf x}_{jk}) \right]
\end{eqnarray*}
and  $T({\bf x})=\infty$ for all points not yet added to the rejection region (i.e., not in $\cup_{h=1}^{j-1} R_h$),
and  $T({\bf x}_{jk'})=j$ if $k'=k$ and $T({\bf x}_{jk'})=\infty$ if $k' \neq k$.
Note, because of the BC conditions (see Section~\ref{sec-Computations}), this is equivalent to $p_U$ as defined in (\ref{eq:Tpvalues}) when $H_0: \theta_2 \geq \theta_1$ (e.g., $b(\theta)=\theta_2-\theta_1$ and $\beta_0=0$).
\item Define $R_j$ as $R_{j-1}$ combined with the point with the lowest value of $p_U({\bf x}_{jk})$, and if there are ties include all tied points, and  define the the associated $T$ function for all included points as $j$.
\end{enumerate}
The top-down CSM ordering is analogous, starts from the highest value of $T({\bf x})$ (i.e.,  $[x_1,x_2] = [0, n_2]$), and uses the other one-sided p-value function, $p_L$.
It is not obvious whether the bottom-up and top-down CSM orderings are equivalent or not.

Barnard's original two-sided CSM ordering is similar, except whenever a point $[x_1^*,x_2^*]$ is included in $R_j$, its symmetric point $[n_1-x_1^*, n_2-x_2^*]$ is also included.

\section*{Acknowledgements}

The authors thank Erica Brittain and anonymous reviewers for comments that improved the paper.

\bibliographystyle{imsart-nameyear}

\bibliography{refs}
\end{document}